\documentclass[submission,copyright,creativecommons,sharealike,noncommercial]{eptcs}
\providecommand{\event}{MSFP 2012}

\usepackage{FormalComparisonGP}

\makeatletter
\@ifundefined{lhs2tex.lhs2tex.sty.read}%
  {\@namedef{lhs2tex.lhs2tex.sty.read}{}%
   \newcommand\SkipToFmtEnd{}%
   \newcommand\EndFmtInput{}%
   \long\def\SkipToFmtEnd#1\EndFmtInput{}%
  }\SkipToFmtEnd

\newcommand\ReadOnlyOnce[1]{\@ifundefined{#1}{\@namedef{#1}{}}\SkipToFmtEnd}
\usepackage{amstext}
\usepackage{amssymb}
\usepackage{stmaryrd}
\DeclareFontFamily{OT1}{cmtex}{}
\DeclareFontShape{OT1}{cmtex}{m}{n}
  {<5><6><7><8>cmtex8
   <9>cmtex9
   <10><10.95><12><14.4><17.28><20.74><24.88>cmtex10}{}
\DeclareFontShape{OT1}{cmtex}{m}{it}
  {<-> ssub * cmtt/m/it}{}

\DeclareFontShape{OT1}{cmtt}{bx}{n}
  {<5><6><7><8>cmtt8
   <9>cmbtt9
   <10><10.95><12><14.4><17.28><20.74><24.88>cmbtt10}{}
\DeclareFontShape{OT1}{cmtex}{bx}{n}
  {<-> ssub * cmtt/bx/n}{}

\newcommand{\Conid}[1]{\mathit{#1}}
\newcommand{\Varid}[1]{\mathit{#1}}
\newcommand{\anonymous}{\kern0.06em \vbox{\hrule\@width.5em}}

\usepackage{polytable}

\@ifundefined{mathindent}%
  {\newdimen\mathindent\mathindent\leftmargini}%
  {}%

\def\resethooks{%
  \global\let\SaveRestoreHook\empty
  \global\let\ColumnHook\empty}
\newcommand*{\savecolumns}[1][default]%
  {\g@addto@macro\SaveRestoreHook{\savecolumns[#1]}}
\newcommand*{\restorecolumns}[1][default]%
  {\g@addto@macro\SaveRestoreHook{\restorecolumns[#1]}}
\newcommand*{\aligncolumn}[2]%
  {\g@addto@macro\ColumnHook{\column{#1}{#2}}}

\resethooks

\newcommand{\onelinecommentchars}{\quad-{}- }
\newcommand{\commentbeginchars}{\enskip\{-}
\newcommand{\commentendchars}{-\}\enskip}

\newcommand{\visiblecomments}{%
  \let\onelinecomment=\onelinecommentchars
  \let\commentbegin=\commentbeginchars
  \let\commentend=\commentendchars}

\newcommand{\invisiblecomments}{%
  \let\onelinecomment=\empty
  \let\commentbegin=\empty
  \let\commentend=\empty}

\visiblecomments

\newlength{\blanklineskip}
\newcommand{\hsindent}[1]{\quad}
\let\hspre\empty
\let\hspost\empty

\EndFmtInput
\makeatother
\ReadOnlyOnce{polycode.fmt}%
\makeatletter

\newcommand{\hsnewpar}[1]%
  {{\parskip=0pt\parindent=0pt\par\vskip #1\noindent}}

\newcommand{\hscodestyle}{}

\newcommand{\sethscode}[1]%
  {\expandafter\let\expandafter\hscode\csname #1\endcsname
   \expandafter\let\expandafter\endhscode\csname end#1\endcsname}

  {\par\noindent
   \advance\leftskip\mathindent
   \hscodestyle
   \let\\=\@normalcr
   \let\hspre\(\let\hspost\)%
   \pboxed}%
  {\endpboxed\)%
   \par\noindent
   \ignorespacesafterend}

  {\hsnewpar\abovedisplayskip
   \advance\leftskip\mathindent
   \hscodestyle
   \let\hspre\(\let\hspost\)%
   \pboxed}%
  {\endpboxed%
   \hsnewpar\belowdisplayskip
   \ignorespacesafterend}

  {\hsnewpar\abovedisplayskip
   \advance\leftskip\mathindent
   \hscodestyle
   \let\\=\@normalcr
   \(\pboxed}%
  {\endpboxed\)%
   \hsnewpar\belowdisplayskip
   \ignorespacesafterend}

\newcommand{\plainhs}{\sethscode{plainhscode}}

\plainhs

  {\hsnewpar\abovedisplayskip
   \advance\leftskip\mathindent
   \hscodestyle
   \let\\=\@normalcr
   \(\parray}%
  {\endparray\)%
   \hsnewpar\belowdisplayskip
   \ignorespacesafterend}

  {\parray}{\endparray}

  {\(\parray}{\endparray\)}

\def\codeframewidth{\arrayrulewidth}
\RequirePackage{calc}

  {\parskip=\abovedisplayskip\par\noindent
   \hscodestyle
   \arrayrulewidth=\codeframewidth
   \tabular{@{}|p{\linewidth-2\arraycolsep-2\arrayrulewidth-2pt}|@{}}%
   \hline\framedhslinecorrect\\{-1.5ex}%
   \let\endoflinesave=\\
   \let\\=\@normalcr
   \(\pboxed}%
  {\endpboxed\)%
   \framedhslinecorrect\endoflinesave{.5ex}\hline
   \endtabular
   \parskip=\belowdisplayskip\par\noindent
   \ignorespacesafterend}

\newcommand{\framedhslinecorrect}[2]%
  {#1[#2]}

  {\(\def\column##1##2{}%
   \let\>\undefined\let\<\undefined\let\\\undefined
   \newcommand\>[1][]{}\newcommand\<[1][]{}\newcommand\\[1][]{}%
   \def\fromto##1##2##3{##3}%
   }{\) }%

  {\let\orighscode=\hscode
   \let\origendhscode=\endhscode
   \def\endhscode{\def\hscode{\endgroup\def\@currenvir{hscode}\\}\begingroup}
   \orighscode\def\hscode{\endgroup\def\@currenvir{hscode}}}%
  {\origendhscode
   \global\let\hscode=\orighscode
   \global\let\endhscode=\origendhscode}%

\makeatother
\EndFmtInput
\ReadOnlyOnce{agda.fmt}%

\RequirePackage[T1]{fontenc}
\RequirePackage[utf8x]{inputenc}
\RequirePackage{ucs}
\RequirePackage{amsfonts}

\DeclareUnicodeCharacter{737}{\textsuperscript{l}}
\DeclareUnicodeCharacter{8718}{\ensuremath{\blacksquare}}
\DeclareUnicodeCharacter{8759}{::}
\DeclareUnicodeCharacter{9669}{\ensuremath{\triangleleft}}
\DeclareUnicodeCharacter{8799}{\ensuremath{\stackrel{\scriptscriptstyle ?}{=}}}
\DeclareUnicodeCharacter{10214}{\ensuremath{\llbracket}}
\DeclareUnicodeCharacter{10215}{\ensuremath{\rrbracket}}

\renewcommand\Varid[1]{\mathord{\textsf{#1}}}
\let\Conid\Varid
\newcommand\Keyword[1]{\textsf{\textbf{#1}}}
\EndFmtInput

\DeclareUnicodeCharacter{9655}{\ensuremath{\triangleright}}
\DeclareUnicodeCharacter{9656}{\ensuremath{\blacktriangleright}}
\DeclareUnicodeCharacter{8649}{\ensuremath{\rightrightarrows}}
\DeclareUnicodeCharacter{0955}{\ensuremath{\lambda}}
\DeclareUnicodeCharacter{0956}{\ensuremath{\mu}}
\DeclareUnicodeCharacter{9653}{\ensuremath{\vartriangle}}
\DeclareUnicodeCharacter{9663}{\ensuremath{\triangledown}}
\DeclareUnicodeCharacter{0966}{\ensuremath{\varphi}}
\DeclareUnicodeCharacter{0968}{\ensuremath{\psi}}
\DeclareUnicodeCharacter{0931}{\ensuremath{\Sigma}}
\DeclareUnicodeCharacter{8245}{\textsf{`}}
\DeclareUnicodeCharacter{8242}{\textsf{'}}
\DeclareUnicodeCharacter{8621}{\ensuremath{\leftrightsquigarrow}}
\DeclareUnicodeCharacter{8599}{\ensuremath{\nnearrow}}
\DeclareUnicodeCharacter{8600}{\ensuremath{\ssearrow}}
\DeclareUnicodeCharacter{8261}{\ensuremath{\Lbag}}
\DeclareUnicodeCharacter{8262}{\ensuremath{\Rbag}}

\SetMathAlphabet{\mathsf}{normal}{OT1}{phv}{m}{n}
\DeclareMathAlphabet{\mathsfbf}{OT1}{phv}{bx}{n}

  \definecolor{blue}{named}{black}
  \definecolor{red}{named}{black}
  \definecolor{darkgreen}{named}{black}
  \definecolor{orange}{named}{black}
  \definecolor{green}{named}{black}

\newcommand{\REM}[3]{}
\newcommand{\Pedro}[1]{}
\newcommand{\Andres}[1]{}

\title{A Formal Comparison of Approaches\\ to Datatype-Generic Programming}
\author{Jos\'e Pedro Magalh\~aes\thanks{Funded by the Portuguese
Foundation for Science and Technology (FCT) via the SFRH/BD/35999/2007 grant.
We thank Johan Jeuring and the anonymous reviewers for the helpful feedback.}
\institute{Utrecht University}
\email{jpm@cs.uu.nl}
\and
Andres L\"oh
\institute{Well-Typed LLP}
\email{andres@well-typed.com}
}
\def\titlerunning{A Formal Comparison of Approaches to Datatype-Generic Programming}
\def\authorrunning{Jos\'e Pedro Magalh\~aes \& Andres L\"oh}
\begin{document}
\maketitle

\begin{abstract}
Datatype-generic programming increases program abstraction and reuse by making 
functions operate uniformly across different types.
Many approaches to generic 
programming have been proposed over the years, most of them for Haskell, but 
recently also for dependently typed languages such as Agda. Different approaches 
vary in expressiveness, ease of use, and implementation techniques. 

Some work has been done in comparing the different approaches informally.
However, to our 
knowledge there have been no attempts to formally prove relations between 
different approaches.
%
We thus present a formal comparison of generic programming libraries. We show
how to formalise different approaches in Agda, including a coinductive
representation, and then establish theorems that relate the approaches to
each other.
We provide constructive proofs of inclusion of one approach in another that can
be used to convert between approaches, helping to reduce code duplication 
across different libraries. Our formalisation also helps in providing a clear
picture of the potential of each approach, especially in relating different 
generic views and their expressiveness. 
\end{abstract}

\section{Introduction}
There are many forms of genericity~\cite{Gibbons2007:Datatype}. Of particular
interest to us is \emph{datatype-genericity}: behavior that is generic over the
structure of types. Functional programming languages typically support the 
definition of algebraic datatypes. Due to their algebraic nature, these
datatypes can be conveniently encoded in a sum-of-products structure. Many
functions can be defined to operate on this structure alone; typical examples
are (de)serialisation, traversals, equality, and enumeration.

Datatype-generic programming (from here on referred to simply as generic
programming) approaches have been especially prolific in
Haskell~\cite{haskellbook}. \polyp~\cite{polyp}, now 15 years old, was the first
approach to generic programming in Haskell, implemented as a pre-processor.
Since then, and especially with the advent of advanced type features such as
generalized algebraic datatypes (GADTs) and type families~\cite{OutsideIn},
numerous other approaches have appeared, most of them implemented directly as a
library.
This abundance is caused by the lack of a clearly superior approach; each
approach has its strengths and weaknesses, uses different implementation
mechanisms, a different generic view~\cite{GenericViews} (i.e.\ a different
structural representation of datatypes), or specialises on a
particular task. Their sheer number and variety makes comparisons difficult, and
can make prospective generic programming users struggle even before actually
writing a generic program, since first they have to choose a library that is
adequate to their needs.

Some effort has been made in comparing different approaches to generic
programming from a practical point of
view~\cite{UUCS2008010,HinzeJeuringLoh:2007}, or to classify
approaches~\cite{GP3D}.
While Generic Haskell~\cite{exploringGH} has been formalised in different
attempts~\cite{formalGH, arity}, no formal comparison between modern approaches
has been attempted, leaving a gap in the knowledge of the relationships between
each approach. We argue that this gap should be filled; for starters, a formal
comparison provides a theoretical foundation for understanding different generic
programming approaches and their differences and similarities. However, the
contribution is not merely of theoretical interest, since a comparison can also
provide means for converting between approaches. Ironically, code duplication
across generic programming libraries is evident: the same function can be
nearly identical in different approaches, yet impossible to reuse, due to the
underlying differences in representation. With a formal proof of inclusion
between two approaches a conversion function can be derived, removing the
duplication of generic code.

In this paper we take the initial steps towards a formal comparison of
generic programming approaches:
\begin{itemize}
\item We encode five distinct, yet related generic programming libraries in
the dependently-typed programming language Agda~\cite{norell2009dependently}.

\item We encode one specific Haskell library using recursive codes, by means
of a coinductive formulation. We pay special attention to this approach, as it
also gives rise to more challenging proofs.

\item We show the relations between the approaches, and reason about them.
While the inclusion relations are the expected, the way to convert between
approaches is often far from straightforward, and reveals subtle differences
between the approaches. Each inclusion is evidenced by a conversion function
that brings codes from one universe into another, enabling generic function
reuse across different approaches.

\item Although fully machine-checked (modulo non-termination), our proofs are in
equational reasoning style and resemble handwritten proofs, remaining clear and
elegant.
\end{itemize}

The rest of this paper proceeds as follows: we first introduce the five
approaches we compare by showing their encoding in Agda in \autoref{sec:gplibs}.
We show the inclusion relations between these approaches in
\autoref{sec:comparison}, and focus on the details of some particular proofs.
Finally we conclude in \autoref{sec:discussion}, discussing shortcomings of our
work and providing directions for future research.

\section{Generic programming libraries}
\label{sec:gplibs}%
In this section we introduce each of the five libraries that we model.
We model all the libraries in Agda, and reduce them to their essence, namely
to the generic view they encode. We leave out details such as implementation
mechanisms (e.g.\ through type classes or GADTs), encoding of meta-information
such as constructor names (it is generally an issue orthogonal to the library),
or modularity.

All five libraries we choose use a sum-of-products representation of data.
One does not use fixed points, while the other four use each a different form of
fixed-point operator. For the latter we show how to define a mapping
function, which can be used to define all the standard recursion
morphisms~\cite{bananas}, and are also necessary for the conversion proofs.
We leave the comparison of libraries using a view
other than sum-of-products for future work (\autoref{sec:discussion}).
Nonetheless, the libraries we choose are rather different in their
expressiveness, especially regarding support for parametrised datatypes and
families of mutually recursive types, as we will show.

\subsection{Regular}
\regular is a simple generic programming library, originally written to support
a generic rewriting system~\cite{1411321}. It has a fixed-point
view on data: the generic representation is a pattern-functor, and a fixed-point
operator ties the recursion explicitly. In the original formulation, this is
used to ensure that rewriting meta-variables can only occur at recursive
positions of the datatype.

We model every library by defining a \ensuremath{\ty{Code}} type that represents the generic
universe, and its interpretation function \ensuremath{\id{⟦\_⟧}} that maps the codes into
Agda types. The universe, interpretation, and fixed-point operator for \regular
follow:

\noindent\begin{minipage}[t]{0.5\linewidth}
\begin{hscode}\SaveRestoreHook
\column{B}{@{}>{\hspre}l<{\hspost}@{}}%
\column{3}{@{}>{\hspre}l<{\hspost}@{}}%
\column{5}{@{}>{\hspre}l<{\hspost}@{}}%
\column{10}{@{}>{\hspre}l<{\hspost}@{}}%
\column{28}{@{}>{\hspre}l<{\hspost}@{}}%
\column{E}{@{}>{\hspre}l<{\hspost}@{}}%
\>[3]{}\Keyword{data}\;\ty{Code}\;\mathbin{:}\;\ty{Set}\;\Keyword{where}{}\<[E]%
\\
\>[3]{}\hsindent{2}{}\<[5]%
\>[5]{}\con{U}\;{}\<[10]%
\>[10]{}\mathbin{:}\;{}\<[28]%
\>[28]{}\ty{Code}{}\<[E]%
\\
\>[3]{}\hsindent{2}{}\<[5]%
\>[5]{}\con{I}\;{}\<[10]%
\>[10]{}\mathbin{:}\;{}\<[28]%
\>[28]{}\ty{Code}{}\<[E]%
\\
\>[3]{}\hsindent{2}{}\<[5]%
\>[5]{}\consymop{⊕}\;{}\<[10]%
\>[10]{}\mathbin{:}\;(\Conid{F}\;\Conid{G}\;\mathbin{:}\;\ty{Code})\;\Varid{→}\;{}\<[28]%
\>[28]{}\ty{Code}{}\<[E]%
\\
\>[3]{}\hsindent{2}{}\<[5]%
\>[5]{}\consymop{⊗}\;{}\<[10]%
\>[10]{}\mathbin{:}\;(\Conid{F}\;\Conid{G}\;\mathbin{:}\;\ty{Code})\;\Varid{→}\;{}\<[28]%
\>[28]{}\ty{Code}{}\<[E]%
\ColumnHook
\end{hscode}\resethooks
\end{minipage}
\begin{minipage}[t]{0.5\linewidth}
\begin{hscode}\SaveRestoreHook
\column{B}{@{}>{\hspre}l<{\hspost}@{}}%
\column{3}{@{}>{\hspre}l<{\hspost}@{}}%
\column{12}{@{}>{\hspre}l<{\hspost}@{}}%
\column{15}{@{}>{\hspre}l<{\hspost}@{}}%
\column{28}{@{}>{\hspre}l<{\hspost}@{}}%
\column{31}{@{}>{\hspre}l<{\hspost}@{}}%
\column{E}{@{}>{\hspre}l<{\hspost}@{}}%
\>[3]{}\id{⟦\_⟧}\;\mathbin{:}\;\ty{Code}\;\Varid{→}\;(\ty{Set}\;\Varid{→}\;\ty{Set}){}\<[E]%
\\
\>[3]{}\id{⟦}\;\con{U}\;{}\<[12]%
\>[12]{}\id{⟧}\;{}\<[15]%
\>[15]{}\Conid{A}\;\mathrel{=}\;\ty{⊤}{}\<[E]%
\\
\>[3]{}\id{⟦}\;\con{I}\;{}\<[12]%
\>[12]{}\id{⟧}\;{}\<[15]%
\>[15]{}\Conid{A}\;\mathrel{=}\;\Conid{A}{}\<[E]%
\\
\>[3]{}\id{⟦}\;\Conid{F}\;\consym{⊕}\;\Conid{G}\;{}\<[12]%
\>[12]{}\id{⟧}\;{}\<[15]%
\>[15]{}\Conid{A}\;\mathrel{=}\;\id{⟦}\;\Conid{F}\;\id{⟧}\;\Conid{A}\;{}\<[28]%
\>[28]{}\ty{⊎}\;{}\<[31]%
\>[31]{}\id{⟦}\;\Conid{G}\;\id{⟧}\;\Conid{A}{}\<[E]%
\\
\>[3]{}\id{⟦}\;\Conid{F}\;\consym{⊗}\;\Conid{G}\;{}\<[12]%
\>[12]{}\id{⟧}\;{}\<[15]%
\>[15]{}\Conid{A}\;\mathrel{=}\;\id{⟦}\;\Conid{F}\;\id{⟧}\;\Conid{A}\;{}\<[28]%
\>[28]{}\ty{×}\;{}\<[31]%
\>[31]{}\id{⟦}\;\Conid{G}\;\id{⟧}\;\Conid{A}{}\<[E]%
\ColumnHook
\end{hscode}\resethooks
\end{minipage}
\vspace{ -1em}
\begin{hscode}\SaveRestoreHook
\column{B}{@{}>{\hspre}l<{\hspost}@{}}%
\column{3}{@{}>{\hspre}l<{\hspost}@{}}%
\column{E}{@{}>{\hspre}l<{\hspost}@{}}%
\>[3]{}\Keyword{data}\;\ty{μ}\;(\Conid{F}\;\mathbin{:}\;\ty{Code})\;\mathbin{:}\;\ty{Set}\;\Keyword{where}\;\con{⟨\_⟩}\;\mathbin{:}\;\id{⟦}\;\Conid{F}\;\id{⟧}\;(\ty{μ}\;\Conid{F})\;\Varid{→}\;\ty{μ}\;\Conid{F}{}\<[E]%
\ColumnHook
\end{hscode}\resethooks
We have codes for units, sums, products, and recursive
positions, denoted by \ensuremath{\con{I}}. The interpretation of unit, sum, and product relies
on the Agda types for unit (\ensuremath{\ty{⊤}}), disjoint sum \ensuremath{(\ty{\ensuremath{\_\!⊎\!\_}})}, and non-dependent
product \ensuremath{(\ty{\ensuremath{\_\!×\!\_}})}, respectively. The interpretation is parametrised over a \ensuremath{\ty{Set}}
that is returned in the \ensuremath{\con{I}} case.

Libraries with a fixed-point view on data allow defining a \ensuremath{\id{map}} function.
In \regular, this function lifts a transformation between sets \ensuremath{\Conid{A}} and \ensuremath{\Conid{B}} to a
transformation between interpretations parametrised over \ensuremath{\Conid{A}} and \ensuremath{\Conid{B}},
simply by applying the function in the \ensuremath{\con{I}} case:

\begin{hscode}\SaveRestoreHook
\column{B}{@{}>{\hspre}l<{\hspost}@{}}%
\column{3}{@{}>{\hspre}l<{\hspost}@{}}%
\column{8}{@{}>{\hspre}l<{\hspost}@{}}%
\column{11}{@{}>{\hspre}l<{\hspost}@{}}%
\column{16}{@{}>{\hspre}l<{\hspost}@{}}%
\column{25}{@{}>{\hspre}l<{\hspost}@{}}%
\column{31}{@{}>{\hspre}l<{\hspost}@{}}%
\column{39}{@{}>{\hspre}l<{\hspost}@{}}%
\column{47}{@{}>{\hspre}l<{\hspost}@{}}%
\column{E}{@{}>{\hspre}l<{\hspost}@{}}%
\>[3]{}\id{map}\;{}\<[8]%
\>[8]{}\mathbin{:}\;{}\<[11]%
\>[11]{}(\Conid{F}\;\mathbin{:}\;\ty{Code})\;\Varid{→}\;\{\mskip1.5mu \Conid{A}\;\Conid{B}\;\mathbin{:}\;\ty{Set}\mskip1.5mu\}\;\Varid{→}\;(\Conid{A}\;\Varid{→}\;\Conid{B})\;\Varid{→}\;\id{⟦}\;\Conid{F}\;\id{⟧}\;\Conid{A}\;\Varid{→}\;\id{⟦}\;\Conid{F}\;\id{⟧}\;\Conid{B}{}\<[E]%
\\[\blanklineskip]%
\>[3]{}\id{map}\;\con{U}\;{}\<[16]%
\>[16]{}\Varid{f}\;\__{}\;{}\<[31]%
\>[31]{}\mathrel{=}\;\con{tt}{}\<[E]%
\\
\>[3]{}\id{map}\;\con{I}\;{}\<[16]%
\>[16]{}\Varid{f}\;\Varid{x}\;{}\<[31]%
\>[31]{}\mathrel{=}\;\Varid{f}\;\Varid{x}{}\<[E]%
\\
\>[3]{}\id{map}\;(\Conid{F}\;\consym{⊕}\;\Conid{G})\;{}\<[16]%
\>[16]{}\Varid{f}\;(\con{inj₁}\;{}\<[25]%
\>[25]{}\Varid{x})\;{}\<[31]%
\>[31]{}\mathrel{=}\;\con{inj₁}\;{}\<[39]%
\>[39]{}(\id{map}\;\Conid{F}\;{}\<[47]%
\>[47]{}\Varid{f}\;\Varid{x}){}\<[E]%
\\
\>[3]{}\id{map}\;(\Conid{F}\;\consym{⊕}\;\Conid{G})\;{}\<[16]%
\>[16]{}\Varid{f}\;(\con{inj₂}\;{}\<[25]%
\>[25]{}\Varid{x})\;{}\<[31]%
\>[31]{}\mathrel{=}\;\con{inj₂}\;{}\<[39]%
\>[39]{}(\id{map}\;\Conid{G}\;{}\<[47]%
\>[47]{}\Varid{f}\;\Varid{x}){}\<[E]%
\\
\>[3]{}\id{map}\;(\Conid{F}\;\consym{⊗}\;\Conid{G})\;{}\<[16]%
\>[16]{}\Varid{f}\;(\Varid{x}\;\consym{,}\,\;\Varid{y})\;{}\<[31]%
\>[31]{}\mathrel{=}\;\id{map}\;\Conid{F}\;\Varid{f}\;\Varid{x}\;\consym{,}\,\;\id{map}\;\Conid{G}\;\Varid{f}\;\Varid{y}{}\<[E]%
\ColumnHook
\end{hscode}\resethooks
The \ensuremath{\id{map}} function can be used to define many other useful generic functions,
most notably recursion morphisms~\cite{bananas}. For example, catamorphisms
are defined as follows:
\begin{hscode}\SaveRestoreHook
\column{B}{@{}>{\hspre}l<{\hspost}@{}}%
\column{3}{@{}>{\hspre}l<{\hspost}@{}}%
\column{E}{@{}>{\hspre}l<{\hspost}@{}}%
\>[3]{}\id{cata}\;\mathbin{:}\;\{\mskip1.5mu \Conid{A}\;\mathbin{:}\;\ty{Set}\mskip1.5mu\}\;\to \;(\Conid{F}\;\mathbin{:}\;\ty{Code})\;\to \;(\id{⟦}\;\Conid{F}\;\id{⟧}\;\Conid{A}\;\to \;\Conid{A})\;\to \;\ty{μ}\;\Conid{F}\;\to \;\Conid{A}{}\<[E]%
\\
\>[3]{}\id{cata}\;\Conid{C}\;\Varid{f}\;\con{⟨}\;\Varid{x}\;\con{⟩}\;\mathrel{=}\;\Varid{f}\;(\id{map}\;\Conid{C}\;(\id{cata}\;\Conid{C}\;\Varid{f})\;\Varid{x}){}\<[E]%
\ColumnHook
\end{hscode}\resethooks
\Pedro{Doesn't pass the termination checker.}%

Datatypes can be encoded by giving their code, such as \ensuremath{\Conid{NatC}} for the natural
numbers, and then taking the fixed point. Hence, a natural number is a value of
type \ensuremath{\ty{μ}\;\Conid{NatC}}; in the example below, \ensuremath{\Varid{aNat}} encodes the number \ensuremath{\Varid{2}}:

\begin{hscode}\SaveRestoreHook
\column{B}{@{}>{\hspre}l<{\hspost}@{}}%
\column{3}{@{}>{\hspre}l<{\hspost}@{}}%
\column{E}{@{}>{\hspre}l<{\hspost}@{}}%
\>[3]{}\Conid{NatC}\;\mathbin{:}\;\ty{Code}{}\<[E]%
\\
\>[3]{}\Conid{NatC}\;\mathrel{=}\;\con{U}\;\consym{⊕}\;\con{I}{}\<[E]%
\\[\blanklineskip]%
\>[3]{}\Varid{aNat}\;\mathbin{:}\;\ty{μ}\;\Conid{NatC}{}\<[E]%
\\
\>[3]{}\Varid{aNat}\;\mathrel{=}\;\con{⟨}\;\con{inj₂}\;\con{⟨}\;\con{inj₂}\;\con{⟨}\;\con{inj₁}\;\con{tt}\;\con{⟩}\;\con{⟩}\;\con{⟩}{}\<[E]%
\ColumnHook
\end{hscode}\resethooks

\subsection{PolyP}
\polyp~\cite{polyp} is an early pre-processor approach to generic programming.
However, this aspect is not essential to \polyp, as it still works with an 
underlying view of Haskell datatypes. This view is very similar to that of
\regular, only that it also abstracts over one datatype
parameter, in addition to one recursive occurrence. \polyp therefore represents
types as bifunctors, whereas \regular uses plain functors. The encoding of
\polyp's universe in Agda follows:

\noindent\begin{minipage}[t]{0.5\linewidth}
\begin{hscode}\SaveRestoreHook
\column{B}{@{}>{\hspre}l<{\hspost}@{}}%
\column{3}{@{}>{\hspre}l<{\hspost}@{}}%
\column{5}{@{}>{\hspre}l<{\hspost}@{}}%
\column{10}{@{}>{\hspre}l<{\hspost}@{}}%
\column{28}{@{}>{\hspre}l<{\hspost}@{}}%
\column{E}{@{}>{\hspre}l<{\hspost}@{}}%
\>[3]{}\Keyword{data}\;\ty{Code}\;\mathbin{:}\;\ty{Set}\;\Keyword{where}{}\<[E]%
\\
\>[3]{}\hsindent{2}{}\<[5]%
\>[5]{}\con{U}\;{}\<[10]%
\>[10]{}\mathbin{:}\;{}\<[28]%
\>[28]{}\ty{Code}{}\<[E]%
\\
\>[3]{}\hsindent{2}{}\<[5]%
\>[5]{}\con{P}\;{}\<[10]%
\>[10]{}\mathbin{:}\;{}\<[28]%
\>[28]{}\ty{Code}{}\<[E]%
\\
\>[3]{}\hsindent{2}{}\<[5]%
\>[5]{}\con{I}\;{}\<[10]%
\>[10]{}\mathbin{:}\;{}\<[28]%
\>[28]{}\ty{Code}{}\<[E]%
\\
\>[3]{}\hsindent{2}{}\<[5]%
\>[5]{}\consymop{⊕}\;{}\<[10]%
\>[10]{}\mathbin{:}\;(\Conid{F}\;\Conid{G}\;\mathbin{:}\;\ty{Code})\;\Varid{→}\;{}\<[28]%
\>[28]{}\ty{Code}{}\<[E]%
\\
\>[3]{}\hsindent{2}{}\<[5]%
\>[5]{}\consymop{⊗}\;{}\<[10]%
\>[10]{}\mathbin{:}\;(\Conid{F}\;\Conid{G}\;\mathbin{:}\;\ty{Code})\;\Varid{→}\;{}\<[28]%
\>[28]{}\ty{Code}{}\<[E]%
\\
\>[3]{}\hsindent{2}{}\<[5]%
\>[5]{}\consymop{⊚}\;{}\<[10]%
\>[10]{}\mathbin{:}\;(\Conid{F}\;\Conid{G}\;\mathbin{:}\;\ty{Code})\;\Varid{→}\;{}\<[28]%
\>[28]{}\ty{Code}{}\<[E]%
\ColumnHook
\end{hscode}\resethooks
\end{minipage}
\begin{minipage}[t]{0.5\linewidth}
\begin{hscode}\SaveRestoreHook
\column{B}{@{}>{\hspre}l<{\hspost}@{}}%
\column{3}{@{}>{\hspre}l<{\hspost}@{}}%
\column{12}{@{}>{\hspre}l<{\hspost}@{}}%
\column{15}{@{}>{\hspre}l<{\hspost}@{}}%
\column{34}{@{}>{\hspre}l<{\hspost}@{}}%
\column{E}{@{}>{\hspre}l<{\hspost}@{}}%
\>[3]{}\id{⟦\_⟧}\;\mathbin{:}\;\ty{Code}\;\Varid{→}\;(\ty{Set}\;\Varid{→}\;\ty{Set}\;\Varid{→}\;\ty{Set}){}\<[E]%
\\
\>[3]{}\id{⟦}\;\con{U}\;{}\<[12]%
\>[12]{}\id{⟧}\;{}\<[15]%
\>[15]{}\Conid{A}\;\Conid{R}\;\mathrel{=}\;\ty{⊤}{}\<[E]%
\\
\>[3]{}\id{⟦}\;\con{P}\;{}\<[12]%
\>[12]{}\id{⟧}\;{}\<[15]%
\>[15]{}\Conid{A}\;\Conid{R}\;\mathrel{=}\;\Conid{A}{}\<[E]%
\\
\>[3]{}\id{⟦}\;\con{I}\;{}\<[12]%
\>[12]{}\id{⟧}\;{}\<[15]%
\>[15]{}\Conid{A}\;\Conid{R}\;\mathrel{=}\;\Conid{R}{}\<[E]%
\\
\>[3]{}\id{⟦}\;\Conid{F}\;\consym{⊕}\;\Conid{G}\;{}\<[12]%
\>[12]{}\id{⟧}\;{}\<[15]%
\>[15]{}\Conid{A}\;\Conid{R}\;\mathrel{=}\;\id{⟦}\;\Conid{F}\;\id{⟧}\;\Conid{A}\;\Conid{R}\;\ty{⊎}\;{}\<[34]%
\>[34]{}\id{⟦}\;\Conid{G}\;\id{⟧}\;\Conid{A}\;\Conid{R}{}\<[E]%
\\
\>[3]{}\id{⟦}\;\Conid{F}\;\consym{⊗}\;\Conid{G}\;{}\<[12]%
\>[12]{}\id{⟧}\;{}\<[15]%
\>[15]{}\Conid{A}\;\Conid{R}\;\mathrel{=}\;\id{⟦}\;\Conid{F}\;\id{⟧}\;\Conid{A}\;\Conid{R}\;\ty{×}\;{}\<[34]%
\>[34]{}\id{⟦}\;\Conid{G}\;\id{⟧}\;\Conid{A}\;\Conid{R}{}\<[E]%
\\
\>[3]{}\id{⟦}\;\Conid{F}\;\consym{⊚}\;\Conid{G}\;{}\<[12]%
\>[12]{}\id{⟧}\;{}\<[15]%
\>[15]{}\Conid{A}\;\Conid{R}\;\mathrel{=}\;\ty{μ}\;\Conid{F}\;(\id{⟦}\;\Conid{G}\;\id{⟧}\;\Conid{A}\;\Conid{R}){}\<[E]%
\ColumnHook
\end{hscode}\resethooks
\end{minipage}
\vspace{ -1em}
\begin{hscode}\SaveRestoreHook
\column{B}{@{}>{\hspre}l<{\hspost}@{}}%
\column{3}{@{}>{\hspre}l<{\hspost}@{}}%
\column{E}{@{}>{\hspre}l<{\hspost}@{}}%
\>[3]{}\Keyword{data}\;\ty{μ}\;(\Conid{F}\;\mathbin{:}\;\ty{Code})\;(\Conid{A}\;\mathbin{:}\;\ty{Set})\;\mathbin{:}\;\ty{Set}\;\Keyword{where}\;\con{⟨\_⟩}\;\mathbin{:}\;\id{⟦}\;\Conid{F}\;\id{⟧}\;\Conid{A}\;(\ty{μ}\;\Conid{F}\;\Conid{A})\;\Varid{→}\;\ty{μ}\;\Conid{F}\;\Conid{A}{}\<[E]%
\ColumnHook
\end{hscode}\resethooks
In the codes, the only differences from \regular are the addition of a \ensuremath{\con{P}}
code, for the parameter, and a code \ensuremath{\consymop{⊚}} for composition.
The interpretation is parametrised over two \ensuremath{\ty{Set}}s, one for the parameter and
the other for the recursive position. Composition is interpreted by taking the
fixed-point of the left bifunctor, thereby closing its recursion, and replacing
its parameters by the interpretation of the right bifunctor. There is at least
one other plausible interpretation for composition, namely interpreting the
left bifunctor with closed right bifunctors as parameter (\ensuremath{\id{⟦}\;\Conid{F}\;\id{⟧}\;(\ty{μ}\;\Conid{G}\;\Conid{A})\;\Conid{R}}),
but we give the interpretation taken by the original implementation.

This asymmetric treatment of the parameters to composition is worth a detailed
discussion. In \polyp, the left functor \ensuremath{\Conid{F}} is first closed under recursion,
and its parameter is set to be the interpretation of the right functor \ensuremath{\Conid{G}}.
The parameter \ensuremath{\Conid{A}} is used in the interpretation of \ensuremath{\Conid{G}}, as is the recursive
position \ensuremath{\Conid{R}}. Care must be taken when using composition to keep in mind the
way it is interpreted. For instance, if we have a code for binary trees with
elements at the leaves \ensuremath{\Conid{TreeC}}, and a code for lists \ensuremath{\Conid{ListC}}, one might naively
think that the code for trees with lists at the leaves is \ensuremath{\Conid{TreeC}\;\consym{⊚}\;\Conid{ListC}}, but
that is not the case. Instead, the
code we are after is \ensuremath{(\Conid{ListC}\;\consym{⊚}\;\Conid{P})\;\consym{⊕}\;(\Conid{I}\;\consym{⊗}\;\Conid{I})}. Apart from requiring careful
usage, this composition does not allow us to reuse the code for trees when
defining trees with lists (although the resulting code quite resembles that of
trees). The \indexed approach (described in
\autoref{sec:indexed}) has a more convenient interpretation of composition;
this subtle difference is revealed explicitly in our conversion from \polyp to
\indexed (\autoref{sec:PolyP2Indexed}).

The fixed-point operator takes a bifunctor and produces a functor, by closing
the recursive positions and leaving the parameter open. The \ensuremath{\id{map}} operation
for bifunctors takes two argument functions, one to apply to parameters, and
the other to apply to recursive positions:
\begin{hscode}\SaveRestoreHook
\column{B}{@{}>{\hspre}l<{\hspost}@{}}%
\column{3}{@{}>{\hspre}l<{\hspost}@{}}%
\column{8}{@{}>{\hspre}l<{\hspost}@{}}%
\column{11}{@{}>{\hspre}l<{\hspost}@{}}%
\column{16}{@{}>{\hspre}l<{\hspost}@{}}%
\column{27}{@{}>{\hspre}l<{\hspost}@{}}%
\column{33}{@{}>{\hspre}l<{\hspost}@{}}%
\column{36}{@{}>{\hspre}l<{\hspost}@{}}%
\column{38}{@{}>{\hspre}l<{\hspost}@{}}%
\column{41}{@{}>{\hspre}l<{\hspost}@{}}%
\column{49}{@{}>{\hspre}l<{\hspost}@{}}%
\column{E}{@{}>{\hspre}l<{\hspost}@{}}%
\>[3]{}\id{map}\;{}\<[8]%
\>[8]{}\mathbin{:}\;{}\<[11]%
\>[11]{}\{\mskip1.5mu \Conid{A}\;\Conid{B}\;\Conid{R}\;\Conid{S}\;\mathbin{:}\;\ty{Set}\mskip1.5mu\}\;(\Conid{F}\;\mathbin{:}\;\ty{Code})\;\Varid{→}\;(\Conid{A}\;\Varid{→}\;\Conid{B})\;\Varid{→}\;(\Conid{R}\;\Varid{→}\;\Conid{S})\;\Varid{→}\;\id{⟦}\;\Conid{F}\;\id{⟧}\;\Conid{A}\;\Conid{R}\;\Varid{→}\;\id{⟦}\;\Conid{F}\;\id{⟧}\;\Conid{B}\;\Conid{S}{}\<[E]%
\\
\>[3]{}\id{map}\;\con{U}\;{}\<[16]%
\>[16]{}\Varid{f}\;\Varid{g}\;\__{}\;{}\<[33]%
\>[33]{}\mathrel{=}\;\con{tt}{}\<[E]%
\\
\>[3]{}\id{map}\;\con{P}\;{}\<[16]%
\>[16]{}\Varid{f}\;\Varid{g}\;\Varid{x}\;{}\<[33]%
\>[33]{}\mathrel{=}\;\Varid{f}\;{}\<[38]%
\>[38]{}\Varid{x}{}\<[E]%
\\
\>[3]{}\id{map}\;\con{I}\;{}\<[16]%
\>[16]{}\Varid{f}\;\Varid{g}\;\Varid{x}\;{}\<[33]%
\>[33]{}\mathrel{=}\;\Varid{g}\;{}\<[38]%
\>[38]{}\Varid{x}{}\<[E]%
\\
\>[3]{}\id{map}\;(\Conid{F}\;\consym{⊕}\;\Conid{G})\;{}\<[16]%
\>[16]{}\Varid{f}\;\Varid{g}\;(\con{inj₁}\;{}\<[27]%
\>[27]{}\Varid{x})\;{}\<[33]%
\>[33]{}\mathrel{=}\;\con{inj₁}\;{}\<[41]%
\>[41]{}(\id{map}\;\Conid{F}\;{}\<[49]%
\>[49]{}\Varid{f}\;\Varid{g}\;\Varid{x}){}\<[E]%
\\
\>[3]{}\id{map}\;(\Conid{F}\;\consym{⊕}\;\Conid{G})\;{}\<[16]%
\>[16]{}\Varid{f}\;\Varid{g}\;(\con{inj₂}\;{}\<[27]%
\>[27]{}\Varid{x})\;{}\<[33]%
\>[33]{}\mathrel{=}\;\con{inj₂}\;{}\<[41]%
\>[41]{}(\id{map}\;\Conid{G}\;{}\<[49]%
\>[49]{}\Varid{f}\;\Varid{g}\;\Varid{x}){}\<[E]%
\\
\>[3]{}\id{map}\;(\Conid{F}\;\consym{⊗}\;\Conid{G})\;{}\<[16]%
\>[16]{}\Varid{f}\;\Varid{g}\;(\Varid{x}\;\consym{,}\,\;\Varid{y})\;{}\<[33]%
\>[33]{}\mathrel{=}\;\id{map}\;\Conid{F}\;\Varid{f}\;\Varid{g}\;\Varid{x}\;\consym{,}\,\;\id{map}\;\Conid{G}\;\Varid{f}\;\Varid{g}\;\Varid{y}{}\<[E]%
\\
\>[3]{}\id{map}\;(\Conid{F}\;\consym{⊚}\;\Conid{G})\;{}\<[16]%
\>[16]{}\Varid{f}\;\Varid{g}\;\con{⟨}\;\Varid{x}\;\con{⟩}\;{}\<[33]%
\>[33]{}\mathrel{=}\;{}\<[36]%
\>[36]{}\con{⟨}\;\id{map}\;\Conid{F}\;(\id{map}\;\Conid{G}\;\Varid{f}\;\Varid{g})\;(\id{map}\;(\Conid{F}\;\consym{⊚}\;\Conid{G})\;\Varid{f}\;\Varid{g})\;\Varid{x}\;\con{⟩}{}\<[E]%
\ColumnHook
\end{hscode}\resethooks

A map over the parameters, \ensuremath{\Varid{pmap}}, operating on fixed points of bifunctors,
can be built from \ensuremath{\id{map}} trivially:
\begin{hscode}\SaveRestoreHook
\column{B}{@{}>{\hspre}l<{\hspost}@{}}%
\column{3}{@{}>{\hspre}l<{\hspost}@{}}%
\column{9}{@{}>{\hspre}l<{\hspost}@{}}%
\column{12}{@{}>{\hspre}l<{\hspost}@{}}%
\column{38}{@{}>{\hspre}l<{\hspost}@{}}%
\column{E}{@{}>{\hspre}l<{\hspost}@{}}%
\>[3]{}\Varid{pmap}\;{}\<[9]%
\>[9]{}\mathbin{:}\;{}\<[12]%
\>[12]{}\{\mskip1.5mu \Conid{A}\;\Conid{B}\;\mathbin{:}\;\ty{Set}\mskip1.5mu\}\;(\Conid{F}\;\mathbin{:}\;\ty{Code})\;\Varid{→}\;{}\<[38]%
\>[38]{}(\Conid{A}\;\Varid{→}\;\Conid{B})\;\Varid{→}\;\ty{μ}\;\Conid{F}\;\Conid{A}\;\Varid{→}\;\ty{μ}\;\Conid{F}\;\Conid{B}{}\<[E]%
\\
\>[3]{}\Varid{pmap}\;\Conid{F}\;\Varid{f}\;\con{⟨}\;\Varid{x}\;\con{⟩}\;\mathrel{=}\;\con{⟨}\;\id{map}\;\Conid{F}\;\Varid{f}\;(\Varid{pmap}\;\Conid{F}\;\Varid{f})\;\Varid{x}\;\con{⟩}{}\<[E]%
\ColumnHook
\end{hscode}\resethooks

As an example encoding in \polyp we show the type of non-empty rose trees:

\vspace*{ -1em}
\noindent\begin{minipage}[t]{0.35\linewidth}
\begin{hscode}\SaveRestoreHook
\column{B}{@{}>{\hspre}l<{\hspost}@{}}%
\column{E}{@{}>{\hspre}l<{\hspost}@{}}%
\>[B]{}\Conid{ListC}\;\mathbin{:}\;\ty{Code}{}\<[E]%
\\
\>[B]{}\Conid{ListC}\;\mathrel{=}\;\con{U}\;\consym{⊕}\;(\con{P}\;\consym{⊗}\;\con{I}){}\<[E]%
\\[\blanklineskip]%
\>[B]{}\Conid{RoseC}\;\mathbin{:}\;\ty{Code}{}\<[E]%
\\
\>[B]{}\Conid{RoseC}\;\mathrel{=}\;\con{P}\;\consym{⊗}\;(\Conid{ListC}\;\consym{⊚}\;\con{I}){}\<[E]%
\ColumnHook
\end{hscode}\resethooks
\end{minipage}
\begin{minipage}[t]{0.65\linewidth}
\begin{hscode}\SaveRestoreHook
\column{B}{@{}>{\hspre}l<{\hspost}@{}}%
\column{E}{@{}>{\hspre}l<{\hspost}@{}}%
\>[B]{}\Varid{sRose}\;\mathbin{:}\;\ty{μ}\;\Conid{RoseC}\;\ty{⊤}{}\<[E]%
\\
\>[B]{}\Varid{sRose}\;\mathrel{=}\;\con{⟨}\;\con{tt}\consym{,}\,\con{⟨}\;\con{inj₁}\;\con{tt}\;\con{⟩}\;\con{⟩}{}\<[E]%
\\[\blanklineskip]%
\>[B]{}\Varid{lRose}\;\mathbin{:}\;\ty{μ}\;\Conid{RoseC}\;\ty{⊤}{}\<[E]%
\\
\>[B]{}\Varid{lRose}\;\mathrel{=}\;\con{⟨}\;\con{tt}\consym{,}\,\con{⟨}\;\con{inj₂}\;(\Varid{sRose}\consym{,}\,\con{⟨}\;\con{inj₂}\;(\Varid{sRose}\consym{,}\,\con{⟨}\;\con{inj₁}\;\con{tt}\;\con{⟩})\;\con{⟩})\;\con{⟩}\;\con{⟩}{}\<[E]%
\ColumnHook
\end{hscode}\resethooks
\end{minipage}
We first encode lists in \ensuremath{\Conid{ListC}}; rose trees are a parameter and a list
containing more rose trees (\ensuremath{\Conid{RoseC}}). The smallest possible rose tree
is \ensuremath{\Varid{sRose}}, containing a single element and an empty list. A larger tree \ensuremath{\Varid{lRose}}
contains a parameter and a list with two small rose trees.

\subsection{Multirec}

The \multirec library~\cite{multirec} is also a generalisation of \regular,
allowing for multiple recursive positions instead of only one. This means that
families of mutually recursive datatypes can be encoded in \multirec. For this,
types are represented as higher-order (or indexed) functors:

\begin{hscode}\SaveRestoreHook
\column{B}{@{}>{\hspre}l<{\hspost}@{}}%
\column{3}{@{}>{\hspre}l<{\hspost}@{}}%
\column{E}{@{}>{\hspre}l<{\hspost}@{}}%
\>[3]{}\id{Indexed}\;\mathbin{:}\;\ty{Set}\;\Varid{→}\;\ty{Set₁}{}\<[E]%
\\
\>[3]{}\id{Indexed}\;\Conid{I}\;\mathrel{=}\;\Conid{I}\;\Varid{→}\;\ty{Set}{}\<[E]%
\ColumnHook
\end{hscode}\resethooks

Codes themselves are parametrised over an index \ensuremath{\ty{Set}}, that is
used in the \ensuremath{\con{I}} case. Furthermore, we have a new code
\ensuremath{\con{!}} for tagging a code with a particular index. The interpretation is
parametrised by a function \ensuremath{\Varid{r}} that maps indices (recursive positions) to
\ensuremath{\ty{Set}}s, and a specific index \ensuremath{\Varid{i}} that defines which particular position we are
interested in; since a code can define several types, the interpretation is a
function from the index of a particular type to its interpretation.
For an occurrence of an index \ensuremath{\con{I}}, we retrieve the associated set using the
function \ensuremath{\Varid{r}}. Tagging constrains the interpretation to a particular index \ensuremath{\Varid{j}},
so we check if \ensuremath{\Varid{j}} is the same as \ensuremath{\Varid{i}}:

\noindent\begin{minipage}[t]{0.43\linewidth}
\begin{hscode}\SaveRestoreHook
\column{B}{@{}>{\hspre}l<{\hspost}@{}}%
\column{3}{@{}>{\hspre}l<{\hspost}@{}}%
\column{5}{@{}>{\hspre}l<{\hspost}@{}}%
\column{10}{@{}>{\hspre}l<{\hspost}@{}}%
\column{28}{@{}>{\hspre}l<{\hspost}@{}}%
\column{31}{@{}>{\hspre}l<{\hspost}@{}}%
\column{E}{@{}>{\hspre}l<{\hspost}@{}}%
\>[3]{}\Keyword{data}\;\ty{Code}\;(\Conid{I}\;\mathbin{:}\;\ty{Set})\;\mathbin{:}\;\ty{Set}\;\Keyword{where}{}\<[E]%
\\
\>[3]{}\hsindent{2}{}\<[5]%
\>[5]{}\con{U}\;{}\<[10]%
\>[10]{}\mathbin{:}\;{}\<[31]%
\>[31]{}\ty{Code}\;\Conid{I}{}\<[E]%
\\
\>[3]{}\hsindent{2}{}\<[5]%
\>[5]{}\con{I}\;{}\<[10]%
\>[10]{}\mathbin{:}\;\Conid{I}\;{}\<[28]%
\>[28]{}\Varid{→}\;{}\<[31]%
\>[31]{}\ty{Code}\;\Conid{I}{}\<[E]%
\\
\>[3]{}\hsindent{2}{}\<[5]%
\>[5]{}\con{!}\;{}\<[10]%
\>[10]{}\mathbin{:}\;\Conid{I}\;{}\<[28]%
\>[28]{}\Varid{→}\;{}\<[31]%
\>[31]{}\ty{Code}\;\Conid{I}{}\<[E]%
\\
\>[3]{}\hsindent{2}{}\<[5]%
\>[5]{}\consymop{⊕}\;{}\<[10]%
\>[10]{}\mathbin{:}\;(\Conid{F}\;\Conid{G}\;\mathbin{:}\;\ty{Code}\;\Conid{I})\;{}\<[28]%
\>[28]{}\Varid{→}\;{}\<[31]%
\>[31]{}\ty{Code}\;\Conid{I}{}\<[E]%
\\
\>[3]{}\hsindent{2}{}\<[5]%
\>[5]{}\consymop{⊗}\;{}\<[10]%
\>[10]{}\mathbin{:}\;(\Conid{F}\;\Conid{G}\;\mathbin{:}\;\ty{Code}\;\Conid{I})\;{}\<[28]%
\>[28]{}\Varid{→}\;{}\<[31]%
\>[31]{}\ty{Code}\;\Conid{I}{}\<[E]%
\ColumnHook
\end{hscode}\resethooks
\end{minipage}
\begin{minipage}[t]{0.57\linewidth}
\begin{hscode}\SaveRestoreHook
\column{B}{@{}>{\hspre}l<{\hspost}@{}}%
\column{3}{@{}>{\hspre}l<{\hspost}@{}}%
\column{12}{@{}>{\hspre}l<{\hspost}@{}}%
\column{15}{@{}>{\hspre}l<{\hspost}@{}}%
\column{32}{@{}>{\hspre}l<{\hspost}@{}}%
\column{35}{@{}>{\hspre}l<{\hspost}@{}}%
\column{E}{@{}>{\hspre}l<{\hspost}@{}}%
\>[3]{}\id{⟦\_⟧}\;\mathbin{:}\;\{\mskip1.5mu \Conid{I}\;\mathbin{:}\;\ty{Set}\mskip1.5mu\}\;\Varid{→}\;\ty{Code}\;\Conid{I}\;\Varid{→}\;\id{Indexed}\;\Conid{I}\;\Varid{→}\;\id{Indexed}\;\Conid{I}{}\<[E]%
\\
\>[3]{}\id{⟦}\;\con{U}\;{}\<[12]%
\>[12]{}\id{⟧}\;{}\<[15]%
\>[15]{}\Varid{r}\;\Varid{i}\;\mathrel{=}\;\ty{⊤}{}\<[E]%
\\
\>[3]{}\id{⟦}\;\con{I}\;\Varid{j}\;{}\<[12]%
\>[12]{}\id{⟧}\;{}\<[15]%
\>[15]{}\Varid{r}\;\Varid{i}\;\mathrel{=}\;\Varid{r}\;\Varid{j}{}\<[E]%
\\
\>[3]{}\id{⟦}\;\con{!}\;\Varid{j}\;{}\<[12]%
\>[12]{}\id{⟧}\;{}\<[15]%
\>[15]{}\Varid{r}\;\Varid{i}\;\mathrel{=}\;\Varid{i}\;\ty{≡}\;\Varid{j}{}\<[E]%
\\
\>[3]{}\id{⟦}\;\Conid{F}\;\consym{⊕}\;\Conid{G}\;{}\<[12]%
\>[12]{}\id{⟧}\;{}\<[15]%
\>[15]{}\Varid{r}\;\Varid{i}\;\mathrel{=}\;\id{⟦}\;\Conid{F}\;\id{⟧}\;\Varid{r}\;\Varid{i}\;{}\<[32]%
\>[32]{}\ty{⊎}\;{}\<[35]%
\>[35]{}\id{⟦}\;\Conid{G}\;\id{⟧}\;\Varid{r}\;\Varid{i}{}\<[E]%
\\
\>[3]{}\id{⟦}\;\Conid{F}\;\consym{⊗}\;\Conid{G}\;{}\<[12]%
\>[12]{}\id{⟧}\;{}\<[15]%
\>[15]{}\Varid{r}\;\Varid{i}\;\mathrel{=}\;\id{⟦}\;\Conid{F}\;\id{⟧}\;\Varid{r}\;\Varid{i}\;{}\<[32]%
\>[32]{}\ty{×}\;{}\<[35]%
\>[35]{}\id{⟦}\;\Conid{G}\;\id{⟧}\;\Varid{r}\;\Varid{i}{}\<[E]%
\ColumnHook
\end{hscode}\resethooks
\end{minipage}

Mapping is entirely similar to the \regular \ensuremath{\id{map}}, only that the function being
mapped is now an index-preserving map. Similarly, the fixed-point operator is
also indexed:
\begin{hscode}\SaveRestoreHook
\column{B}{@{}>{\hspre}l<{\hspost}@{}}%
\column{3}{@{}>{\hspre}l<{\hspost}@{}}%
\column{E}{@{}>{\hspre}l<{\hspost}@{}}%
\>[3]{}\Keyword{data}\;\ty{μ}\;\{\mskip1.5mu \Conid{I}\;\mathbin{:}\;\ty{Set}\mskip1.5mu\}\;(\Conid{F}\;\mathbin{:}\;\ty{Code}\;\Conid{I})\;(\Varid{i}\;\mathbin{:}\;\Conid{I})\;\mathbin{:}\;\ty{Set}\;\Keyword{where}\;\con{⟨\_⟩}\;\mathbin{:}\;\id{⟦}\;\Conid{F}\;\id{⟧}\;(\ty{μ}\;\Conid{F})\;\Varid{i}\;\Varid{→}\;\ty{μ}\;\Conid{F}\;\Varid{i}{}\<[E]%
\ColumnHook
\end{hscode}\resethooks
\vspace{ -2em}
\begin{hscode}\SaveRestoreHook
\column{B}{@{}>{\hspre}l<{\hspost}@{}}%
\column{3}{@{}>{\hspre}l<{\hspost}@{}}%
\column{8}{@{}>{\hspre}l<{\hspost}@{}}%
\column{11}{@{}>{\hspre}l<{\hspost}@{}}%
\column{16}{@{}>{\hspre}l<{\hspost}@{}}%
\column{27}{@{}>{\hspre}l<{\hspost}@{}}%
\column{33}{@{}>{\hspre}l<{\hspost}@{}}%
\column{41}{@{}>{\hspre}l<{\hspost}@{}}%
\column{49}{@{}>{\hspre}l<{\hspost}@{}}%
\column{E}{@{}>{\hspre}l<{\hspost}@{}}%
\>[3]{}\id{map}\;{}\<[8]%
\>[8]{}\mathbin{:}\;{}\<[11]%
\>[11]{}\{\mskip1.5mu \Conid{I}\;\mathbin{:}\;\ty{Set}\mskip1.5mu\}\;\{\mskip1.5mu \Conid{R}\;\Conid{S}\;\mathbin{:}\;\id{Indexed}\;\Conid{I}\mskip1.5mu\}\;(\Conid{F}\;\mathbin{:}\;\ty{Code}\;\Conid{I})\;{}\<[E]%
\\
\>[8]{}\Varid{→}\;{}\<[11]%
\>[11]{}(\Varid{∀}\;\Varid{i}\;\Varid{→}\;\Conid{R}\;\Varid{i}\;\Varid{→}\;\Conid{S}\;\Varid{i})\;\Varid{→}\;(\Varid{∀}\;\Varid{i}\;\Varid{→}\;\id{⟦}\;\Conid{F}\;\id{⟧}\;\Conid{R}\;\Varid{i}\;\Varid{→}\;\id{⟦}\;\Conid{F}\;\id{⟧}\;\Conid{S}\;\Varid{i}){}\<[E]%
\\
\>[3]{}\id{map}\;\con{U}\;{}\<[16]%
\>[16]{}\Varid{f}\;\Varid{i}\;\__{}\;{}\<[33]%
\>[33]{}\mathrel{=}\;\con{tt}{}\<[E]%
\\
\>[3]{}\id{map}\;(\con{I}\;\Varid{j})\;{}\<[16]%
\>[16]{}\Varid{f}\;\Varid{i}\;\Varid{x}\;{}\<[33]%
\>[33]{}\mathrel{=}\;\Varid{f}\;\Varid{j}\;\Varid{x}{}\<[E]%
\\
\>[3]{}\id{map}\;(\con{!}\;\Varid{j})\;{}\<[16]%
\>[16]{}\Varid{f}\;\Varid{i}\;\Varid{x}\;{}\<[33]%
\>[33]{}\mathrel{=}\;\Varid{x}{}\<[E]%
\\
\>[3]{}\id{map}\;(\Conid{F}\;\consym{⊕}\;\Conid{G})\;{}\<[16]%
\>[16]{}\Varid{f}\;\Varid{i}\;(\con{inj₁}\;{}\<[27]%
\>[27]{}\Varid{x})\;{}\<[33]%
\>[33]{}\mathrel{=}\;\con{inj₁}\;{}\<[41]%
\>[41]{}(\id{map}\;\Conid{F}\;{}\<[49]%
\>[49]{}\Varid{f}\;\Varid{i}\;\Varid{x}){}\<[E]%
\\
\>[3]{}\id{map}\;(\Conid{F}\;\consym{⊕}\;\Conid{G})\;{}\<[16]%
\>[16]{}\Varid{f}\;\Varid{i}\;(\con{inj₂}\;{}\<[27]%
\>[27]{}\Varid{x})\;{}\<[33]%
\>[33]{}\mathrel{=}\;\con{inj₂}\;{}\<[41]%
\>[41]{}(\id{map}\;\Conid{G}\;{}\<[49]%
\>[49]{}\Varid{f}\;\Varid{i}\;\Varid{x}){}\<[E]%
\\
\>[3]{}\id{map}\;(\Conid{F}\;\consym{⊗}\;\Conid{G})\;{}\<[16]%
\>[16]{}\Varid{f}\;\Varid{i}\;(\Varid{x}\;\consym{,}\,\;\Varid{y})\;{}\<[33]%
\>[33]{}\mathrel{=}\;\id{map}\;\Conid{F}\;\Varid{f}\;\Varid{i}\;\Varid{x}\;\consym{,}\,\;\id{map}\;\Conid{G}\;\Varid{f}\;\Varid{i}\;\Varid{y}{}\<[E]%
\ColumnHook
\end{hscode}\resethooks

To show an example involving mutually recursive types we encode a zig-zag
sequence of even length. Consider first the family we wish to encode, inside a
\ensuremath{\Keyword{mutual}} block as the datatypes are mutually recursive:
\begin{hscode}\SaveRestoreHook
\column{B}{@{}>{\hspre}l<{\hspost}@{}}%
\column{3}{@{}>{\hspre}l<{\hspost}@{}}%
\column{5}{@{}>{\hspre}l<{\hspost}@{}}%
\column{7}{@{}>{\hspre}l<{\hspost}@{}}%
\column{12}{@{}>{\hspre}l<{\hspost}@{}}%
\column{17}{@{}>{\hspre}l<{\hspost}@{}}%
\column{E}{@{}>{\hspre}l<{\hspost}@{}}%
\>[3]{}\Keyword{mutual}{}\<[E]%
\\
\>[3]{}\hsindent{2}{}\<[5]%
\>[5]{}\Keyword{data}\;\ty{Zig}\;\mathbin{:}\;{}\<[17]%
\>[17]{}\ty{Set}\;\Keyword{where}{}\<[E]%
\\
\>[5]{}\hsindent{2}{}\<[7]%
\>[7]{}\con{zig}\;{}\<[12]%
\>[12]{}\mathbin{:}\;\ty{Zag}\;\to \;\ty{Zig}{}\<[E]%
\\
\>[5]{}\hsindent{2}{}\<[7]%
\>[7]{}\con{end}\;{}\<[12]%
\>[12]{}\mathbin{:}\;\ty{Zig}{}\<[E]%
\\[\blanklineskip]%
\>[3]{}\hsindent{2}{}\<[5]%
\>[5]{}\Keyword{data}\;\ty{Zag}\;\mathbin{:}\;{}\<[17]%
\>[17]{}\ty{Set}\;\Keyword{where}{}\<[E]%
\\
\>[5]{}\hsindent{2}{}\<[7]%
\>[7]{}\con{zag}\;\mathbin{:}\;\ty{Zig}\;\to \;\ty{Zag}{}\<[E]%
\ColumnHook
\end{hscode}\resethooks
We can encode this family in \multirec as follows:
\begin{hscode}\SaveRestoreHook
\column{B}{@{}>{\hspre}l<{\hspost}@{}}%
\column{3}{@{}>{\hspre}l<{\hspost}@{}}%
\column{9}{@{}>{\hspre}l<{\hspost}@{}}%
\column{11}{@{}>{\hspre}l<{\hspost}@{}}%
\column{E}{@{}>{\hspre}l<{\hspost}@{}}%
\>[3]{}\Conid{ZigC}\;{}\<[9]%
\>[9]{}\mathbin{:}\;\ty{Code}\;(\ty{⊤}\;\ty{⊎}\;\ty{⊤}){}\<[E]%
\\
\>[3]{}\Conid{ZigC}\;{}\<[11]%
\>[11]{}\mathrel{=}\;\con{I}\;(\con{inj₂}\;\con{tt})\;\consym{⊕}\;\con{U}{}\<[E]%
\\[\blanklineskip]%
\>[3]{}\Conid{ZagC}\;{}\<[9]%
\>[9]{}\mathbin{:}\;\ty{Code}\;(\ty{⊤}\;\ty{⊎}\;\ty{⊤}){}\<[E]%
\\
\>[3]{}\Conid{ZagC}\;{}\<[11]%
\>[11]{}\mathrel{=}\;\con{I}\;(\con{inj₁}\;\con{tt}){}\<[E]%
\\[\blanklineskip]%
\>[3]{}\Conid{ZigZagC}\;\mathbin{:}\;\ty{Code}\;(\ty{⊤}\;\ty{⊎}\;\ty{⊤}){}\<[E]%
\\
\>[3]{}\Conid{ZigZagC}\;\mathrel{=}\;(\con{!}\;(\con{inj₁}\;\con{tt})\;\consym{⊗}\;\Conid{ZigC})\;\consym{⊕}\;(\con{!}\;(\con{inj₂}\;\con{tt})\;\consym{⊗}\;\Conid{ZagC}){}\<[E]%
\\[\blanklineskip]%
\>[3]{}\Varid{zigZagEnd}\;\mathbin{:}\;\ty{μ}\;\Conid{ZigZagC}\;(\con{inj₁}\;\con{tt}){}\<[E]%
\\
\>[3]{}\Varid{zigZagEnd}\;\mathrel{=}\;\con{⟨}\;\con{inj₁}\;(\con{refl}\consym{,}\,\con{inj₁}\;\con{⟨}\;\con{inj₂}\;(\con{refl}\consym{,}\,\con{⟨}\;\con{inj₁}\;(\con{refl}\consym{,}\,\con{inj₂}\;\con{tt})\;\con{⟩})\;\con{⟩})\;\con{⟩}{}\<[E]%
\ColumnHook
\end{hscode}\resethooks
\ensuremath{\Varid{zigZagEnd}} encodes the value \ensuremath{\con{zig}\;(\con{zag}\;\con{end})}, as its name suggests. Note how we 
define the code for each type in the family separately (\ensuremath{\Conid{ZigC}} and \ensuremath{\Conid{ZagC}}), and 
then a code \ensuremath{\Conid{ZigZagC}} for the family, encoding a tagged choice between the two 
types. As a consequence, proofs of index equality (witnessed by \ensuremath{\con{refl}}) are 
present throughout the encoded values.

\subsection{Indexed Functors}
\label{sec:indexed}%

Just like \multirec can be seen as a generalisation of \regular to multiple
recursive positions, the \indexed approach~\cite{jpm:gpif:11} can be seen as a
generalisation
of \polyp both to multiple recursive positions and multiple parameters. In
\indexed, datatypes are represented by functors which are not only indexed on
their input (like \multirec) but also on their output. It is related to other
approaches to dependently-typed generic programming, such as the levitating
universe of Epigram 2~\cite{levitation}, or the theory of
indexed containers~\cite{indexedcontainers}.

The added complexity makes \indexed cumbersome to encode in Haskell, so its
original description was in Agda (although recent developments in GHC's kind
system~\cite{jpm:ghp:12} might now allow us to write a Haskell version of this
library). Below we show a subset of its universe; we elide the reindexing,
sigma, and isomorphism operators from the original presentation:

\noindent\begin{minipage}[c]{0.5\linewidth}
\begin{hscode}\SaveRestoreHook
\column{B}{@{}>{\hspre}l<{\hspost}@{}}%
\column{3}{@{}>{\hspre}l<{\hspost}@{}}%
\column{E}{@{}>{\hspre}l<{\hspost}@{}}%
\>[3]{}\id{Indexed}\;\mathbin{:}\;\ty{Set}\;\Varid{→}\;\ty{Set₁}{}\<[E]%
\\
\>[3]{}\id{Indexed}\;\Conid{I}\;\mathrel{=}\;\Conid{I}\;\Varid{→}\;\ty{Set}{}\<[E]%
\ColumnHook
\end{hscode}\resethooks
\end{minipage}
\begin{minipage}[c]{0.5\linewidth}
\begin{hscode}\SaveRestoreHook
\column{B}{@{}>{\hspre}l<{\hspost}@{}}%
\column{3}{@{}>{\hspre}l<{\hspost}@{}}%
\column{8}{@{}>{\hspre}l<{\hspost}@{}}%
\column{11}{@{}>{\hspre}l<{\hspost}@{}}%
\column{18}{@{}>{\hspre}l<{\hspost}@{}}%
\column{26}{@{}>{\hspre}l<{\hspost}@{}}%
\column{E}{@{}>{\hspre}l<{\hspost}@{}}%
\>[3]{}\id{\ensuremath{\_\!∣\!\_}}\;{}\<[8]%
\>[8]{}\mathbin{:}\;{}\<[11]%
\>[11]{}\Varid{∀}\;\{\mskip1.5mu \Conid{I}\;\Conid{J}\mskip1.5mu\}\;\Varid{→}\;\id{Indexed}\;\Conid{I}\;\Varid{→}\;\id{Indexed}\;\Conid{J}\;{}\<[E]%
\\
\>[8]{}\Varid{→}\;{}\<[11]%
\>[11]{}\id{Indexed}\;(\Conid{I}\;\ty{⊎}\;\Conid{J}){}\<[E]%
\\
\>[3]{}(\Varid{r}\;\idsym{∣}\;\Varid{s})\;(\con{inj₁}\;{}\<[18]%
\>[18]{}\Varid{i})\;\mathrel{=}\;\Varid{r}\;{}\<[26]%
\>[26]{}\Varid{i}{}\<[E]%
\\
\>[3]{}(\Varid{r}\;\idsym{∣}\;\Varid{s})\;(\con{inj₂}\;{}\<[18]%
\>[18]{}\Varid{i})\;\mathrel{=}\;\Varid{s}\;{}\<[26]%
\>[26]{}\Varid{i}{}\<[E]%
\ColumnHook
\end{hscode}\resethooks
\end{minipage}

\vspace{ -1em}
\noindent\begin{minipage}[t]{0.5\linewidth}
\begin{hscode}\SaveRestoreHook
\column{B}{@{}>{\hspre}l<{\hspost}@{}}%
\column{3}{@{}>{\hspre}l<{\hspost}@{}}%
\column{5}{@{}>{\hspre}l<{\hspost}@{}}%
\column{11}{@{}>{\hspre}l<{\hspost}@{}}%
\column{14}{@{}>{\hspre}l<{\hspost}@{}}%
\column{17}{@{}>{\hspre}l<{\hspost}@{}}%
\column{20}{@{}>{\hspre}l<{\hspost}@{}}%
\column{32}{@{}>{\hspre}l<{\hspost}@{}}%
\column{E}{@{}>{\hspre}l<{\hspost}@{}}%
\>[3]{}\Keyword{data}\;\ty{Code}\;(\Conid{I}\;\Conid{O}\;\mathbin{:}\;\ty{Set})\;\mathbin{:}\;\ty{Set₁}\;\Keyword{where}{}\<[E]%
\\
\>[3]{}\hsindent{2}{}\<[5]%
\>[5]{}\con{U}\;{}\<[11]%
\>[11]{}\mathbin{:}\;{}\<[20]%
\>[20]{}\ty{Code}\;\Conid{I}\;\Conid{O}{}\<[E]%
\\
\>[3]{}\hsindent{2}{}\<[5]%
\>[5]{}\con{I}\;{}\<[11]%
\>[11]{}\mathbin{:}\;{}\<[14]%
\>[14]{}\Conid{I}\;{}\<[17]%
\>[17]{}\Varid{→}\;{}\<[20]%
\>[20]{}\ty{Code}\;\Conid{I}\;\Conid{O}{}\<[E]%
\\
\>[3]{}\hsindent{2}{}\<[5]%
\>[5]{}\con{!}\;{}\<[11]%
\>[11]{}\mathbin{:}\;{}\<[14]%
\>[14]{}\Conid{O}\;{}\<[17]%
\>[17]{}\Varid{→}\;{}\<[20]%
\>[20]{}\ty{Code}\;\Conid{I}\;\Conid{O}{}\<[E]%
\\
\>[3]{}\hsindent{2}{}\<[5]%
\>[5]{}\consymop{⊕}\;{}\<[11]%
\>[11]{}\mathbin{:}\;{}\<[14]%
\>[14]{}(\Conid{F}\;\Conid{G}\;\mathbin{:}\;\ty{Code}\;\Conid{I}\;\Conid{O})\;{}\<[32]%
\>[32]{}\Varid{→}\;\ty{Code}\;\Conid{I}\;\Conid{O}{}\<[E]%
\\
\>[3]{}\hsindent{2}{}\<[5]%
\>[5]{}\consymop{⊗}\;{}\<[11]%
\>[11]{}\mathbin{:}\;{}\<[14]%
\>[14]{}(\Conid{F}\;\Conid{G}\;\mathbin{:}\;\ty{Code}\;\Conid{I}\;\Conid{O})\;{}\<[32]%
\>[32]{}\Varid{→}\;\ty{Code}\;\Conid{I}\;\Conid{O}{}\<[E]%
\\
\>[3]{}\hsindent{2}{}\<[5]%
\>[5]{}\consymop{⊚}\;{}\<[11]%
\>[11]{}\mathbin{:}\;{}\<[14]%
\>[14]{}\{\mskip1.5mu \Conid{M}\;\mathbin{:}\;\ty{Set}\mskip1.5mu\}\;\Varid{→}\;(\Conid{F}\;\mathbin{:}\;\ty{Code}\;\Conid{M}\;\Conid{O})\;{}\<[E]%
\\
\>[11]{}\Varid{→}\;{}\<[14]%
\>[14]{}(\Conid{G}\;\mathbin{:}\;\ty{Code}\;\Conid{I}\;\Conid{M})\;\Varid{→}\;\ty{Code}\;\Conid{I}\;\Conid{O}{}\<[E]%
\\
\>[3]{}\hsindent{2}{}\<[5]%
\>[5]{}\con{Fix}\;{}\<[11]%
\>[11]{}\mathbin{:}\;{}\<[14]%
\>[14]{}(\Conid{F}\;\mathbin{:}\;\ty{Code}\;(\Conid{I}\;\ty{⊎}\;\Conid{O})\;\Conid{O})\;\Varid{→}\;\ty{Code}\;\Conid{I}\;\Conid{O}{}\<[E]%
\ColumnHook
\end{hscode}\resethooks
\end{minipage}
\begin{minipage}[t]{0.5\linewidth}
\begin{hscode}\SaveRestoreHook
\column{B}{@{}>{\hspre}l<{\hspost}@{}}%
\column{3}{@{}>{\hspre}l<{\hspost}@{}}%
\column{8}{@{}>{\hspre}l<{\hspost}@{}}%
\column{11}{@{}>{\hspre}l<{\hspost}@{}}%
\column{12}{@{}>{\hspre}l<{\hspost}@{}}%
\column{15}{@{}>{\hspre}l<{\hspost}@{}}%
\column{22}{@{}>{\hspre}l<{\hspost}@{}}%
\column{33}{@{}>{\hspre}l<{\hspost}@{}}%
\column{36}{@{}>{\hspre}l<{\hspost}@{}}%
\column{E}{@{}>{\hspre}l<{\hspost}@{}}%
\>[3]{}\id{⟦\_⟧}\;{}\<[8]%
\>[8]{}\mathbin{:}\;{}\<[11]%
\>[11]{}\{\mskip1.5mu \Conid{I}\;\Conid{O}\;\mathbin{:}\;\ty{Set}\mskip1.5mu\}\;\Varid{→}\;\ty{Code}\;\Conid{I}\;\Conid{O}\;{}\<[E]%
\\
\>[8]{}\Varid{→}\;{}\<[11]%
\>[11]{}\id{Indexed}\;\Conid{I}\;\Varid{→}\;\id{Indexed}\;\Conid{O}{}\<[E]%
\\
\>[3]{}\id{⟦}\;\con{U}\;{}\<[12]%
\>[12]{}\id{⟧}\;{}\<[15]%
\>[15]{}\Varid{r}\;\Varid{i}\;\mathrel{=}\;{}\<[22]%
\>[22]{}\ty{⊤}{}\<[E]%
\\
\>[3]{}\id{⟦}\;\con{I}\;\Varid{j}\;{}\<[12]%
\>[12]{}\id{⟧}\;{}\<[15]%
\>[15]{}\Varid{r}\;\Varid{i}\;\mathrel{=}\;{}\<[22]%
\>[22]{}\Varid{r}\;\Varid{j}{}\<[E]%
\\
\>[3]{}\id{⟦}\;\con{!}\;\Varid{j}\;{}\<[12]%
\>[12]{}\id{⟧}\;{}\<[15]%
\>[15]{}\Varid{r}\;\Varid{i}\;\mathrel{=}\;{}\<[22]%
\>[22]{}\Varid{i}\;\ty{≡}\;\Varid{j}{}\<[E]%
\\
\>[3]{}\id{⟦}\;\Conid{F}\;\consym{⊕}\;\Conid{G}\;{}\<[12]%
\>[12]{}\id{⟧}\;{}\<[15]%
\>[15]{}\Varid{r}\;\Varid{i}\;\mathrel{=}\;{}\<[22]%
\>[22]{}\id{⟦}\;\Conid{F}\;\id{⟧}\;\Varid{r}\;\Varid{i}\;{}\<[33]%
\>[33]{}\ty{⊎}\;{}\<[36]%
\>[36]{}\id{⟦}\;\Conid{G}\;\id{⟧}\;\Varid{r}\;\Varid{i}{}\<[E]%
\\
\>[3]{}\id{⟦}\;\Conid{F}\;\consym{⊗}\;\Conid{G}\;{}\<[12]%
\>[12]{}\id{⟧}\;{}\<[15]%
\>[15]{}\Varid{r}\;\Varid{i}\;\mathrel{=}\;{}\<[22]%
\>[22]{}\id{⟦}\;\Conid{F}\;\id{⟧}\;\Varid{r}\;\Varid{i}\;{}\<[33]%
\>[33]{}\ty{×}\;{}\<[36]%
\>[36]{}\id{⟦}\;\Conid{G}\;\id{⟧}\;\Varid{r}\;\Varid{i}{}\<[E]%
\\
\>[3]{}\id{⟦}\;\Conid{F}\;\consym{⊚}\;\Conid{G}\;{}\<[12]%
\>[12]{}\id{⟧}\;{}\<[15]%
\>[15]{}\Varid{r}\;\Varid{i}\;\mathrel{=}\;{}\<[22]%
\>[22]{}\id{⟦}\;\Conid{F}\;\id{⟧}\;(\id{⟦}\;\Conid{G}\;\id{⟧}\;\Varid{r})\;\Varid{i}{}\<[E]%
\\
\>[3]{}\id{⟦}\;\con{Fix}\;\Conid{F}\;{}\<[12]%
\>[12]{}\id{⟧}\;{}\<[15]%
\>[15]{}\Varid{r}\;\Varid{i}\;\mathrel{=}\;{}\<[22]%
\>[22]{}\ty{μ}\;\Conid{F}\;\Varid{r}\;\Varid{i}{}\<[E]%
\ColumnHook
\end{hscode}\resethooks
\end{minipage}

\vspace{ -1em}
\begin{hscode}\SaveRestoreHook
\column{B}{@{}>{\hspre}l<{\hspost}@{}}%
\column{3}{@{}>{\hspre}l<{\hspost}@{}}%
\column{5}{@{}>{\hspre}l<{\hspost}@{}}%
\column{E}{@{}>{\hspre}l<{\hspost}@{}}%
\>[3]{}\Keyword{data}\;\ty{μ}\;\{\mskip1.5mu \Conid{I}\;\Conid{O}\;\mathbin{:}\;\ty{Set}\mskip1.5mu\}\;(\Conid{F}\;\mathbin{:}\;\ty{Code}\;(\Conid{I}\;\ty{⊎}\;\Conid{O})\;\Conid{O})\;(\Varid{r}\;\mathbin{:}\;\id{Indexed}\;\Conid{I})\;(\Varid{o}\;\mathbin{:}\;\Conid{O})\;\mathbin{:}\;\ty{Set}\;\Keyword{where}{}\<[E]%
\\
\>[3]{}\hsindent{2}{}\<[5]%
\>[5]{}\con{⟨\_⟩}\;\mathbin{:}\;\id{⟦}\;\Conid{F}\;\id{⟧}\;(\Varid{r}\;\idsym{∣}\;\ty{μ}\;\Conid{F}\;\Varid{r})\;\Varid{o}\;\Varid{→}\;\ty{μ}\;\Conid{F}\;\Varid{r}\;\Varid{o}{}\<[E]%
\ColumnHook
\end{hscode}\resethooks
A major difference from the previous approaches is that the fixed-point
operator is contained within the universe.
Composition is also allowed, and, unlike in \polyp, it does not require
taking a fixed point. Codes are parametrised over two \ensuremath{\ty{Set}}s, which can be
thought of as the input set (parameters) and output set (types defined).
Composition is only possible for codes with composable indices; for instance,
a family of three types with two parameters can be composed with a family
of two types with one parameter, resulting in the original family of
three types, but now taking only one parameter. The fixed-point operator
takes a code with tagged input indices (parameters on the left, recursive
occurrences on the right) and closes the recursive occurrences, producing a
code with only parameters as input.

The \ensuremath{\id{map}} function lifts a transformation \ensuremath{\Varid{r}\;\idsym{⇉}\;\Varid{s}} between indexed functors
\ensuremath{\Varid{r}} and \ensuremath{\Varid{s}} to a transformation \ensuremath{\id{⟦}\;\Conid{F}\;\id{⟧}\;\Varid{r}\;\idsym{⇉}\;\id{⟦}\;\Conid{F}\;\id{⟧}\;\Varid{s}} between interpretations.
In the \ensuremath{\con{Fix}} case, care has to be taken to ensure that the mapping function
\ensuremath{\Varid{f}} is applied to the parameters on the left, and \ensuremath{\id{map}} to the recursive
positions on the right:

\begin{hscode}\SaveRestoreHook
\column{B}{@{}>{\hspre}l<{\hspost}@{}}%
\column{3}{@{}>{\hspre}l<{\hspost}@{}}%
\column{16}{@{}>{\hspre}l<{\hspost}@{}}%
\column{18}{@{}>{\hspre}l<{\hspost}@{}}%
\column{27}{@{}>{\hspre}l<{\hspost}@{}}%
\column{28}{@{}>{\hspre}l<{\hspost}@{}}%
\column{33}{@{}>{\hspre}l<{\hspost}@{}}%
\column{41}{@{}>{\hspre}l<{\hspost}@{}}%
\column{49}{@{}>{\hspre}l<{\hspost}@{}}%
\column{E}{@{}>{\hspre}l<{\hspost}@{}}%
\>[3]{}\id{\ensuremath{\_\!⇉\!\_}}\;\mathbin{:}\;\Varid{∀}\;\{\mskip1.5mu \Conid{I}\mskip1.5mu\}\;\Varid{→}\;(\Conid{R}\;\Conid{S}\;\mathbin{:}\;\id{Indexed}\;\Conid{I})\;\Varid{→}\;\ty{Set}{}\<[E]%
\\
\>[3]{}\Varid{r}\;\idsym{⇉}\;\Varid{s}\;\mathrel{=}\;(\Varid{i}\;\mathbin{:}\;\__{})\;\Varid{→}\;\Varid{r}\;\Varid{i}\;\Varid{→}\;\Varid{s}\;\Varid{i}{}\<[E]%
\\[\blanklineskip]%
\>[3]{}\id{\_∥\_}\;\mathbin{:}\;\Varid{∀}\;\{\mskip1.5mu \Conid{I}\;\Conid{J}\mskip1.5mu\}\;\{\mskip1.5mu \Conid{A}\;\Conid{C}\;\mathbin{:}\;\id{Indexed}\;\Conid{I}\mskip1.5mu\}\;\{\mskip1.5mu \Conid{B}\;\Conid{D}\;\mathbin{:}\;\id{Indexed}\;\Conid{J}\mskip1.5mu\}\;\Varid{→}\;\Conid{A}\;\idsym{⇉}\;\Conid{C}\;\Varid{→}\;\Conid{B}\;\idsym{⇉}\;\Conid{D}\;\Varid{→}\;(\Conid{A}\;\idsym{∣}\;\Conid{B})\;\idsym{⇉}\;(\Conid{C}\;\idsym{∣}\;\Conid{D}){}\<[E]%
\\
\>[3]{}(\Varid{f}\;\idsym{∥}\;\Varid{g})\;(\con{inj₁}\;{}\<[18]%
\>[18]{}\Varid{x})\;\Varid{z}\;\mathrel{=}\;\Varid{f}\;{}\<[28]%
\>[28]{}\Varid{x}\;\Varid{z}{}\<[E]%
\\
\>[3]{}(\Varid{f}\;\idsym{∥}\;\Varid{g})\;(\con{inj₂}\;{}\<[18]%
\>[18]{}\Varid{x})\;\Varid{z}\;\mathrel{=}\;\Varid{g}\;{}\<[28]%
\>[28]{}\Varid{x}\;\Varid{z}{}\<[E]%
\\
\>[3]{}\;{}\<[E]%
\\
\>[3]{}\id{map}\;\mathbin{:}\;\{\mskip1.5mu \Conid{I}\;\Conid{O}\;\mathbin{:}\;\ty{Set}\mskip1.5mu\}\;\{\mskip1.5mu \Varid{r}\;\Varid{s}\;\mathbin{:}\;\id{Indexed}\;\Conid{I}\mskip1.5mu\}\;(\Conid{F}\;\mathbin{:}\;\ty{Code}\;\Conid{I}\;\Conid{O})\;\Varid{→}\;(\Varid{r}\;\idsym{⇉}\;\Varid{s})\;\Varid{→}\;\id{⟦}\;\Conid{F}\;\id{⟧}\;\Varid{r}\;\idsym{⇉}\;\id{⟦}\;\Conid{F}\;\id{⟧}\;\Varid{s}{}\<[E]%
\\
\>[3]{}\id{map}\;\con{U}\;{}\<[16]%
\>[16]{}\Varid{f}\;\Varid{i}\;\__{}\;{}\<[33]%
\>[33]{}\mathrel{=}\;\con{tt}{}\<[E]%
\\
\>[3]{}\id{map}\;(\con{I}\;\Varid{j})\;{}\<[16]%
\>[16]{}\Varid{f}\;\Varid{i}\;\Varid{x}\;{}\<[33]%
\>[33]{}\mathrel{=}\;\Varid{f}\;\Varid{j}\;\Varid{x}{}\<[E]%
\\
\>[3]{}\id{map}\;(\con{!}\;\Varid{j})\;{}\<[16]%
\>[16]{}\Varid{f}\;\Varid{i}\;\Varid{x}\;{}\<[33]%
\>[33]{}\mathrel{=}\;\Varid{x}{}\<[E]%
\\
\>[3]{}\id{map}\;(\Conid{F}\;\consym{⊕}\;\Conid{G})\;{}\<[16]%
\>[16]{}\Varid{f}\;\Varid{i}\;(\con{inj₁}\;{}\<[27]%
\>[27]{}\Varid{x})\;{}\<[33]%
\>[33]{}\mathrel{=}\;\con{inj₁}\;{}\<[41]%
\>[41]{}(\id{map}\;\Conid{F}\;{}\<[49]%
\>[49]{}\Varid{f}\;\Varid{i}\;\Varid{x}){}\<[E]%
\\
\>[3]{}\id{map}\;(\Conid{F}\;\consym{⊕}\;\Conid{G})\;{}\<[16]%
\>[16]{}\Varid{f}\;\Varid{i}\;(\con{inj₂}\;{}\<[27]%
\>[27]{}\Varid{x})\;{}\<[33]%
\>[33]{}\mathrel{=}\;\con{inj₂}\;{}\<[41]%
\>[41]{}(\id{map}\;\Conid{G}\;{}\<[49]%
\>[49]{}\Varid{f}\;\Varid{i}\;\Varid{x}){}\<[E]%
\\
\>[3]{}\id{map}\;(\Conid{F}\;\consym{⊗}\;\Conid{G})\;{}\<[16]%
\>[16]{}\Varid{f}\;\Varid{i}\;(\Varid{x}\;\consym{,}\,\;\Varid{y})\;{}\<[33]%
\>[33]{}\mathrel{=}\;\id{map}\;\Conid{F}\;\Varid{f}\;\Varid{i}\;\Varid{x}\;\consym{,}\,\;\id{map}\;\Conid{G}\;\Varid{f}\;\Varid{i}\;\Varid{y}{}\<[E]%
\\
\>[3]{}\id{map}\;(\Conid{F}\;\consym{⊚}\;\Conid{G})\;{}\<[16]%
\>[16]{}\Varid{f}\;\Varid{i}\;\Varid{x}\;{}\<[33]%
\>[33]{}\mathrel{=}\;\id{map}\;\Conid{F}\;(\id{map}\;\Conid{G}\;\Varid{f})\;\Varid{i}\;\Varid{x}{}\<[E]%
\\
\>[3]{}\id{map}\;(\con{Fix}\;\Conid{F})\;{}\<[16]%
\>[16]{}\Varid{f}\;\Varid{i}\;\con{⟨}\;\Varid{x}\;\con{⟩}\;{}\<[33]%
\>[33]{}\mathrel{=}\;\con{⟨}\;\id{map}\;\Conid{F}\;(\Varid{f}\;\idsym{∥}\;\id{map}\;(\con{Fix}\;\Conid{F})\;\Varid{f})\;\Varid{i}\;\Varid{x}\;\con{⟩}{}\<[E]%
\ColumnHook
\end{hscode}\resethooks
For more information and examples of how to encode datatypes in \indexed,
we refer the reader to the paper that introduced this
approach~\cite{jpm:gpif:11}.

\subsection{Instant Generics}

\ig~\cite{IG} is another approach to generic programming in Haskell with type
families. It distinguishes itself from all the other approaches we have discussed
so far in that it does not represent recursion via a fixed-point combinator.
Like \regular, \ig also
supports a generic rewriting library~\cite{jpm:ladgr:10}. To allow
meta-variables to occur at any position (i.e.\ not only in recursive
positions), type-safe runtime casts are performed to determine if the type of
the meta-variable matches that of the expression.
\ig is also rather similar to the generic programming support recently built
into the Glasgow and Utrecht Haskell compilers~\cite{jpm:gdmh:10}.

In the original encoding of \ig in Haskell, recursive datatypes are handled
through indirect recursion between the conversion functions (between a
datatype and its generic representation) and the generic functions. We find
that the most natural way to model this in Agda is to use
coinduction~\cite{subtyping}. This allows us to define infinite codes, and
generic functions operating on these codes, while still passing the
termination check.
This encoding would also be appropriate for other Haskell approaches without a
fixed-point operator, such as ``Generics for the Masses''~\cite{GM} and
\ligd~\cite{LIGD}. Although approaches without a fixed-point operator have
trouble expressing recursive morphisms, they have been popular in Haskell
because they easily allow encoding datatypes with irregular forms of recursion
(such as mutually recursive or nested~\cite{nested} datatypes).

Compared to the previous approaches, the novelties in the universe of \ig are
a code \ensuremath{\con{K}} for arbitrary \ensuremath{\ty{Set}}s, and a code \ensuremath{\con{R}} for tagging recursive codes.
We give the interpretation as a datatype to ensure that it is inductive%
\footnote{Alternatively, we could use the experimental Agda flag
\text{\tt \char45{}\char45{}guardedness\char45{}preserving\char45{}type\char45{}constructors} to treat type constructors as
inductive constructors when checking productivity.}. The
judicious use of the coinduction primitive \ensuremath{\Varid{♭}} in the \ensuremath{\con{R}} case makes the Agda
encoding pass the termination checker, as the definitions remain productive:

\noindent\begin{minipage}[c]{0.4\linewidth}
\begin{hscode}\SaveRestoreHook
\column{B}{@{}>{\hspre}l<{\hspost}@{}}%
\column{3}{@{}>{\hspre}l<{\hspost}@{}}%
\column{5}{@{}>{\hspre}l<{\hspost}@{}}%
\column{11}{@{}>{\hspre}l<{\hspost}@{}}%
\column{27}{@{}>{\hspre}l<{\hspost}@{}}%
\column{30}{@{}>{\hspre}l<{\hspost}@{}}%
\column{E}{@{}>{\hspre}l<{\hspost}@{}}%
\>[3]{}\Keyword{data}\;\ty{Code}\;\mathbin{:}\;\ty{Set₁}\;\Keyword{where}{}\<[E]%
\\
\>[3]{}\hsindent{2}{}\<[5]%
\>[5]{}\con{U}\;{}\<[11]%
\>[11]{}\mathbin{:}\;{}\<[30]%
\>[30]{}\ty{Code}{}\<[E]%
\\
\>[3]{}\hsindent{2}{}\<[5]%
\>[5]{}\con{K}\;{}\<[11]%
\>[11]{}\mathbin{:}\;\ty{Set}\;{}\<[27]%
\>[27]{}\Varid{→}\;{}\<[30]%
\>[30]{}\ty{Code}{}\<[E]%
\\
\>[3]{}\hsindent{2}{}\<[5]%
\>[5]{}\con{R}\;{}\<[11]%
\>[11]{}\mathbin{:}\;(\Conid{C}\;\mathbin{:}\;\Varid{∞}\;\ty{Code})\;{}\<[27]%
\>[27]{}\Varid{→}\;{}\<[30]%
\>[30]{}\ty{Code}{}\<[E]%
\\
\>[3]{}\hsindent{2}{}\<[5]%
\>[5]{}\consymop{⊕}\;{}\<[11]%
\>[11]{}\mathbin{:}\;(\Conid{C}\;\Conid{D}\;\mathbin{:}\;\ty{Code})\;{}\<[27]%
\>[27]{}\Varid{→}\;{}\<[30]%
\>[30]{}\ty{Code}{}\<[E]%
\\
\>[3]{}\hsindent{2}{}\<[5]%
\>[5]{}\consymop{⊗}\;{}\<[11]%
\>[11]{}\mathbin{:}\;(\Conid{C}\;\Conid{D}\;\mathbin{:}\;\ty{Code})\;{}\<[27]%
\>[27]{}\Varid{→}\;{}\<[30]%
\>[30]{}\ty{Code}{}\<[E]%
\ColumnHook
\end{hscode}\resethooks
\end{minipage}
\begin{minipage}[c]{0.6\linewidth}
\begin{hscode}\SaveRestoreHook
\column{B}{@{}>{\hspre}l<{\hspost}@{}}%
\column{3}{@{}>{\hspre}l<{\hspost}@{}}%
\column{5}{@{}>{\hspre}l<{\hspost}@{}}%
\column{14}{@{}>{\hspre}l<{\hspost}@{}}%
\column{30}{@{}>{\hspre}l<{\hspost}@{}}%
\column{47}{@{}>{\hspre}l<{\hspost}@{}}%
\column{50}{@{}>{\hspre}l<{\hspost}@{}}%
\column{59}{@{}>{\hspre}l<{\hspost}@{}}%
\column{E}{@{}>{\hspre}l<{\hspost}@{}}%
\>[3]{}\Keyword{data}\;\id{⟦\_⟧}\;\mathbin{:}\;\ty{Code}\;\Varid{→}\;\ty{Set₁}\;\Keyword{where}{}\<[E]%
\\
\>[3]{}\hsindent{2}{}\<[5]%
\>[5]{}\con{tt}\;{}\<[14]%
\>[14]{}\mathbin{:}\;{}\<[50]%
\>[50]{}\id{⟦}\;\con{U}\;{}\<[59]%
\>[59]{}\id{⟧}{}\<[E]%
\\
\>[3]{}\hsindent{2}{}\<[5]%
\>[5]{}\con{k}\;{}\<[14]%
\>[14]{}\mathbin{:}\;\{\mskip1.5mu \Conid{A}\;\mathbin{:}\;\ty{Set}\mskip1.5mu\}\;{}\<[30]%
\>[30]{}\Varid{→}\;\Conid{A}\;{}\<[47]%
\>[47]{}\Varid{→}\;{}\<[50]%
\>[50]{}\id{⟦}\;\con{K}\;\Conid{A}\;{}\<[59]%
\>[59]{}\id{⟧}{}\<[E]%
\\
\>[3]{}\hsindent{2}{}\<[5]%
\>[5]{}\con{rec}\;{}\<[14]%
\>[14]{}\mathbin{:}\;\{\mskip1.5mu \Conid{C}\;\mathbin{:}\;\Varid{∞}\;\ty{Code}\mskip1.5mu\}\;{}\<[30]%
\>[30]{}\Varid{→}\;\id{⟦}\;\Varid{♭}\;\Conid{C}\;\id{⟧}\;{}\<[47]%
\>[47]{}\Varid{→}\;{}\<[50]%
\>[50]{}\id{⟦}\;\con{R}\;\Conid{C}\;{}\<[59]%
\>[59]{}\id{⟧}{}\<[E]%
\\
\>[3]{}\hsindent{2}{}\<[5]%
\>[5]{}\con{inj₁}\;{}\<[14]%
\>[14]{}\mathbin{:}\;\{\mskip1.5mu \Conid{C}\;\Conid{D}\;\mathbin{:}\;\ty{Code}\mskip1.5mu\}\;{}\<[30]%
\>[30]{}\Varid{→}\;\id{⟦}\;\Conid{C}\;\id{⟧}\;{}\<[47]%
\>[47]{}\Varid{→}\;{}\<[50]%
\>[50]{}\id{⟦}\;\Conid{C}\;\consym{⊕}\;\Conid{D}\;{}\<[59]%
\>[59]{}\id{⟧}{}\<[E]%
\\
\>[3]{}\hsindent{2}{}\<[5]%
\>[5]{}\con{inj₂}\;{}\<[14]%
\>[14]{}\mathbin{:}\;\{\mskip1.5mu \Conid{C}\;\Conid{D}\;\mathbin{:}\;\ty{Code}\mskip1.5mu\}\;{}\<[30]%
\>[30]{}\Varid{→}\;\id{⟦}\;\Conid{D}\;\id{⟧}\;{}\<[47]%
\>[47]{}\Varid{→}\;{}\<[50]%
\>[50]{}\id{⟦}\;\Conid{C}\;\consym{⊕}\;\Conid{D}\;{}\<[59]%
\>[59]{}\id{⟧}{}\<[E]%
\\
\>[3]{}\hsindent{2}{}\<[5]%
\>[5]{}\consymop{,}\;{}\<[14]%
\>[14]{}\mathbin{:}\;\{\mskip1.5mu \Conid{C}\;\Conid{D}\;\mathbin{:}\;\ty{Code}\mskip1.5mu\}\;{}\<[30]%
\>[30]{}\Varid{→}\;\id{⟦}\;\Conid{C}\;\id{⟧}\;\Varid{→}\;\id{⟦}\;\Conid{D}\;\id{⟧}\;{}\<[47]%
\>[47]{}\Varid{→}\;{}\<[50]%
\>[50]{}\id{⟦}\;\Conid{C}\;\consym{⊗}\;\Conid{D}\;{}\<[59]%
\>[59]{}\id{⟧}{}\<[E]%
\ColumnHook
\end{hscode}\resethooks
\end{minipage}

Note the encoding of lists in \ig:
\begin{hscode}\SaveRestoreHook
\column{B}{@{}>{\hspre}l<{\hspost}@{}}%
\column{3}{@{}>{\hspre}l<{\hspost}@{}}%
\column{E}{@{}>{\hspre}l<{\hspost}@{}}%
\>[3]{}\Conid{ListC}\;\mathbin{:}\;\ty{Set}\;\Varid{→}\;\ty{Code}{}\<[E]%
\\
\>[3]{}\Conid{ListC}\;\Conid{A}\;\mathrel{=}\;\con{U}\;\consym{⊕}\;(\con{K}\;\Conid{A}\;\consym{⊗}\;\con{R}\;(\Varid{♯}\;\Conid{ListC}\;\Conid{A})){}\<[E]%
\ColumnHook
\end{hscode}\resethooks
The definition for \ensuremath{\Conid{ListC}} is directly recursive; since it remains productive,
it is accepted by the termination checker. Due to the lack of fixed points, we
cannot write a map function. But we can easily write other generic functions,
also recursive, such as a traversal that crushes a term into a result:
\begin{hscode}\SaveRestoreHook
\column{B}{@{}>{\hspre}l<{\hspost}@{}}%
\column{3}{@{}>{\hspre}l<{\hspost}@{}}%
\column{18}{@{}>{\hspre}l<{\hspost}@{}}%
\column{49}{@{}>{\hspre}l<{\hspost}@{}}%
\column{64}{@{}>{\hspre}l<{\hspost}@{}}%
\column{E}{@{}>{\hspre}l<{\hspost}@{}}%
\>[3]{}\Varid{crush}\;\mathbin{:}\;\{\mskip1.5mu \Conid{R}\;\mathbin{:}\;\ty{Set}\mskip1.5mu\}\;(\Conid{A}\;\mathbin{:}\;\ty{Code})\;\Varid{→}\;(\Conid{R}\;\Varid{→}\;\Conid{R}\;\Varid{→}\;\Conid{R})\;\Varid{→}\;(\Conid{R}\;\Varid{→}\;\Conid{R})\;\Varid{→}\;\Conid{R}\;\Varid{→}\;\id{⟦}\;\Conid{A}\;\id{⟧}\;\Varid{→}\;\Conid{R}{}\<[E]%
\\[\blanklineskip]%
\>[3]{}\Varid{crush}\;\con{U}\;{}\<[18]%
\>[18]{}\id{\ensuremath{\_\!\boxplus\!\_}}\;\id{\ensuremath{\Uparrow}}\;\id{\ensuremath{\mathds{1}}}\;\__{}\;{}\<[49]%
\>[49]{}\mathrel{=}\;\id{\ensuremath{\mathds{1}}}{}\<[E]%
\\
\>[3]{}\Varid{crush}\;(\con{K}\;\Varid{y})\;{}\<[18]%
\>[18]{}\id{\ensuremath{\_\!\boxplus\!\_}}\;\id{\ensuremath{\Uparrow}}\;\id{\ensuremath{\mathds{1}}}\;\__{}\;{}\<[49]%
\>[49]{}\mathrel{=}\;\id{\ensuremath{\mathds{1}}}{}\<[E]%
\\
\>[3]{}\Varid{crush}\;(\con{R}\;\Conid{C})\;{}\<[18]%
\>[18]{}\id{\ensuremath{\_\!\boxplus\!\_}}\;\id{\ensuremath{\Uparrow}}\;\id{\ensuremath{\mathds{1}}}\;(\con{rec}\;\Varid{x})\;{}\<[49]%
\>[49]{}\mathrel{=}\;\id{\ensuremath{\Uparrow}}\;(\Varid{crush}\;(\Varid{♭}\;\Conid{C})\;\id{\ensuremath{\_\!\boxplus\!\_}}\;\id{\ensuremath{\Uparrow}}\;\id{\ensuremath{\mathds{1}}}\;\Varid{x}){}\<[E]%
\\
\>[3]{}\Varid{crush}\;(\Conid{C}\;\consym{⊕}\;\Conid{D})\;{}\<[18]%
\>[18]{}\id{\ensuremath{\_\!\boxplus\!\_}}\;\id{\ensuremath{\Uparrow}}\;\id{\ensuremath{\mathds{1}}}\;(\con{inj₁}\;\Varid{x})\;{}\<[49]%
\>[49]{}\mathrel{=}\;\Varid{crush}\;\Conid{C}\;{}\<[64]%
\>[64]{}\id{\ensuremath{\_\!\boxplus\!\_}}\;\id{\ensuremath{\Uparrow}}\;\id{\ensuremath{\mathds{1}}}\;\Varid{x}{}\<[E]%
\\
\>[3]{}\Varid{crush}\;(\Conid{C}\;\consym{⊕}\;\Conid{D})\;{}\<[18]%
\>[18]{}\id{\ensuremath{\_\!\boxplus\!\_}}\;\id{\ensuremath{\Uparrow}}\;\id{\ensuremath{\mathds{1}}}\;(\con{inj₂}\;\Varid{x})\;{}\<[49]%
\>[49]{}\mathrel{=}\;\Varid{crush}\;\Conid{D}\;{}\<[64]%
\>[64]{}\id{\ensuremath{\_\!\boxplus\!\_}}\;\id{\ensuremath{\Uparrow}}\;\id{\ensuremath{\mathds{1}}}\;\Varid{x}{}\<[E]%
\\
\>[3]{}\Varid{crush}\;(\Conid{C}\;\consym{⊗}\;\Conid{D})\;{}\<[18]%
\>[18]{}\id{\ensuremath{\_\!\boxplus\!\_}}\;\id{\ensuremath{\Uparrow}}\;\id{\ensuremath{\mathds{1}}}\;(\Varid{x}\;\consym{,}\,\;\Varid{y})\;{}\<[49]%
\>[49]{}\mathrel{=}\;(\Varid{crush}\;\Conid{C}\;\id{\ensuremath{\_\!\boxplus\!\_}}\;\id{\ensuremath{\Uparrow}}\;\id{\ensuremath{\mathds{1}}}\;\Varid{x})\;\id{\ensuremath{\boxplus}}\;(\Varid{crush}\;\Conid{D}\;\id{\ensuremath{\_\!\boxplus\!\_}}\;\id{\ensuremath{\Uparrow}}\;\id{\ensuremath{\mathds{1}}}\;\Varid{y}){}\<[E]%
\ColumnHook
\end{hscode}\resethooks
Function \ensuremath{\Varid{crush}} is similar to \ensuremath{\id{map}} in the sense that it can be used to define
many generic functions. It takes three arguments that specify how to combine
the results of each constructor argument (\ensuremath{\id{\ensuremath{\_\!\boxplus\!\_}}}), how to adapt the result of
a recursive call (\ensuremath{\id{\ensuremath{\Uparrow}}}), and what to return for constants and constructors
with no arguments (\ensuremath{\id{\ensuremath{\mathds{1}}}}).\footnote{It is worth noting that \ensuremath{\Varid{crush}} is itself
a simplified catamorphism for the \ensuremath{\ty{Code}} type.}
However, \ensuremath{\Varid{crush}} is unable to change the type of datatype parameters, since
\ig has no knowledge of parameters.

We can compute the size of a structure as a crush, for instance:
\begin{hscode}\SaveRestoreHook
\column{B}{@{}>{\hspre}l<{\hspost}@{}}%
\column{3}{@{}>{\hspre}l<{\hspost}@{}}%
\column{E}{@{}>{\hspre}l<{\hspost}@{}}%
\>[3]{}\Varid{size}\;\mathbin{:}\;(\Conid{A}\;\mathbin{:}\;\ty{Code})\;\Varid{→}\;\id{⟦}\;\Conid{A}\;\id{⟧}\;\Varid{→}\;\ty{ℕ}{}\<[E]%
\\
\>[3]{}\Varid{size}\;\Conid{C}\;\mathrel{=}\;\Varid{crush}\;\Conid{C}\;\ty{\ensuremath{\_\!+\!\_}}\;\id{suc}\;\Varid{0}{}\<[E]%
\ColumnHook
\end{hscode}\resethooks
Here we combine multiple results by adding them, increment the total at
every recursive call, and ignore constants and units for size purposes.
We can test that this function behaves as expected on lists:
\begin{hscode}\SaveRestoreHook
\column{B}{@{}>{\hspre}l<{\hspost}@{}}%
\column{3}{@{}>{\hspre}l<{\hspost}@{}}%
\column{28}{@{}>{\hspre}l<{\hspost}@{}}%
\column{E}{@{}>{\hspre}l<{\hspost}@{}}%
\>[3]{}\Varid{aList}\;\mathbin{:}\;\id{⟦}\;\Conid{ListC}\;\ty{⊤}\;\id{⟧}{}\<[E]%
\\
\>[3]{}\Varid{aList}\;\mathrel{=}\;\con{inj₂}\;(\con{k}\;\con{tt}\;\consym{,}\,\;\con{rec}\;(\con{inj₂}\;(\con{k}\;\con{tt}\;\consym{,}\,\;\con{rec}\;(\con{inj₁}\;\con{tt})))){}\<[E]%
\\[\blanklineskip]%
\>[3]{}\Varid{testSize}\;\mathbin{:}\;\Varid{size}\;\__{}\;\Varid{aList}\;{}\<[28]%
\>[28]{}\ty{≡}\;\Varid{2}{}\<[E]%
\\
\>[3]{}\Varid{testSize}\;\mathrel{=}\;\con{refl}{}\<[E]%
\ColumnHook
\end{hscode}\resethooks

While a \ensuremath{\id{map}} function cannot be defined like in the
previous approaches, traversal and transformation functions can still be
expressed in \ig. In particular, if one is willing to exchange static by dynamic
type checking, type-safe runtime casts can be performed to compare the types
of elements being mapped against the type expected by the mapping function,
resulting in convenient to use generic functions~\cite{jpm:ladgr:10}. However,
runtime casting is known to result in poor runtime performance, as it prevents
the compiler from performing type-directed optimisations~\cite{jpm:ogie:10}.

\subsection{Summary}
We have shown an Agda encoding for the five libraries we compare. In order to
simplify the proofs for the remainder of the paper, we have omitted a few 
details:
\begin{itemize}
\item In their original formulation, all libraries supported embedding \ensuremath{\ty{Set}}s
into the universe (like the \ensuremath{\con{K}} code in \ig). We omit these for simplicity,
except in \ig where they are essential for embedding datatype parameters.

\item We paid no attention to ease of use of the encodings. Adding
isomorphisms to the universe~\cite{jpm:gpif:11}, for instance, would make the
encodings easier to use in practice. However, for our purposes of formal
modelling, isomorphisms would serve only to enlarge the proofs, with no added
benefit. Therefore we decided to omit them from the model.

\item The variant of \indexed we consider here is significantly simpler than
its original presentation~\cite{jpm:gpif:11}. In particular, we omit the sigma
code, which is used for encoding indexed types. If we were to consider this
code then \indexed would no longer embed fully into \ig
(\autoref{sec:Indexed2IG}), since the latter does not support indexed types.
\end{itemize}

\Pedro{Say something about inductive vs. coinductive fixed points?}%

\section{Comparing the approaches}
\label{sec:comparison}%
\begin{figure}[htb]
\centering{
\begin{tikzpicture}[%
  scale=1.0,
  lib/.style={rectangle,draw=blue!70!black,fill=blue!20!white,thick,minimum width=3.2cm,minimum height=1.7cm,text centered,text width=3.2cm},
  embeds/.style={ ->,very thick}
]%
\node (regular)     at ( 0, 2.5) [lib] {\regular\\\small \ensuremath{\con{I}_{\Varid{1}}}\ \,\ensuremath{\consym{⊕}}\,\ \,\ensuremath{\consym{⊗}}};
\node (multirec)    at (-3.3, 0) [lib] {\multirec\\\small \ensuremath{\con{I}_{\Varid{n}}}\ \,\ensuremath{\consym{⊕}}\,\ \,\ensuremath{\consym{⊗}}};
\node (polyp)       at ( 3.3, 0) [lib] {\polyp\\\small \ensuremath{\con{I}_{\Varid{1}}}\ \ensuremath{\con{P}_{\Varid{1}}}\ \,\ensuremath{\consym{⊕}}\,\ \,\ensuremath{\consym{⊗}}\,\ \,\ensuremath{\consym{⊚}}};
\node (indexed)     at ( 0,-2.5) [lib] {\indexed\\\small \ensuremath{\con{I}_{\Varid{n}}}\ \ensuremath{\con{P}_{\Varid{n}}} \,\ensuremath{\consym{⊕}}\,\ \,\ensuremath{\consym{⊗}}\,\ \,\ensuremath{\consym{⊚}}\,\ \ensuremath{\con{Fix}}};
\node (coinductive) at ( 5,-2.5) [lib] {\ig\\\small \ensuremath{\con{R}} \,\ensuremath{\consym{⊕}}\,\ \,\ensuremath{\consym{⊗}}};
\draw [embeds] (regular)  -- (polyp);
\draw [embeds] (indexed)  -- (coinductive);
\draw [embeds] (multirec) -- (indexed);
\draw [embeds] (regular)  -- (multirec);
\draw [embeds] (polyp)    -- (indexed);
\end{tikzpicture}}
\caption{Embedding relation between the approaches}
\label{fig:diagram}
\end{figure}
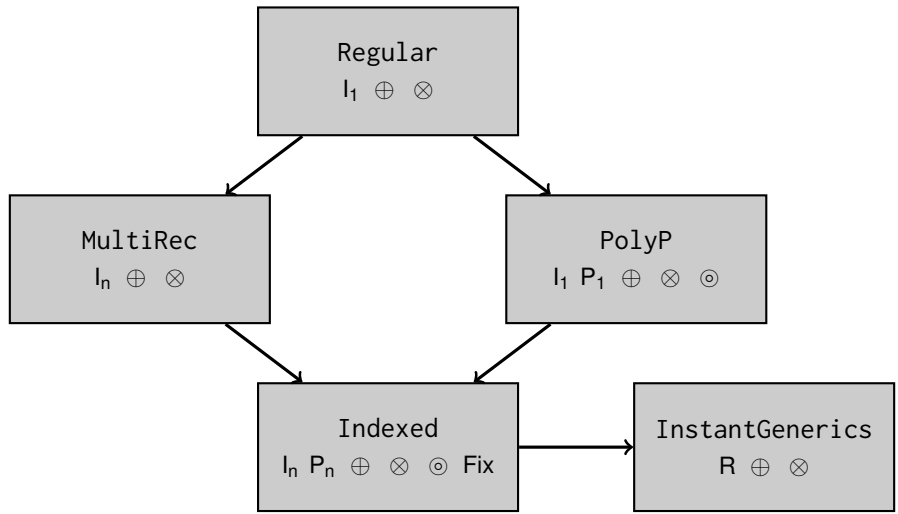

We now proceed to describe how the approaches relate to each other. We show
which approaches can be \emph{embedded} in other approaches; when we say that
approach \text{\tt A} embeds into approach \text{\tt B}, we mean that the interpretation of any
code defined in approach \text{\tt A} has an equivalent interpretation in approach \text{\tt B}.
The starting point of an embedding is a code-conversion function that maps codes
from approach \text{\tt A} into approach \text{\tt B}. \autoref{fig:diagram} presents a graphical
view of the embedding relation between the five approaches; the arrows mean
``embeds into''. Note that the embedding relation is naturally transitive.
\Pedro{But we haven't shown this.}%
As expected, \multirec and \polyp both subsume \regular, but they don't
subsume each other, since one supports families of recursive types and the
other supports one parameter. They are however both subsumed by \indexed.
Finally, the liberal encoding of \ig allows encoding at least all the types
supported by the other approaches (even if it doesn't support the same
operations on those types, such as catamorphisms).

We will now focus on some of the conversions and their proofs, choosing those
that are particularly interesting.

\subsection{Regular to PolyP}
\label{sec:Regular2PolyP}%
We start with the simplest relation: embedding \regular into \polyp. The first
step is to convert \regular codes into \polyp codes:

\begin{hscode}\SaveRestoreHook
\column{B}{@{}>{\hspre}l<{\hspost}@{}}%
\column{3}{@{}>{\hspre}l<{\hspost}@{}}%
\column{27}{@{}>{\hspre}l<{\hspost}@{}}%
\column{E}{@{}>{\hspre}l<{\hspost}@{}}%
\>[3]{}{}_{\Varid{r}}\!\!\!\Uparrow^{_{\Varid{p}}}\!\!\;\mathbin{:}\;\ty{Code}_{\Varid{r}}\;\Varid{→}\;\ty{Code}_{\Varid{p}}{}\<[E]%
\\
\>[3]{}{}_{\Varid{r}}\!\!\!\Uparrow^{_{\Varid{p}}}\!\!\;\con{U}_{\Varid{r}}\;{}\<[27]%
\>[27]{}\mathrel{=}\;\con{U}_{\Varid{p}}{}\<[E]%
\\
\>[3]{}{}_{\Varid{r}}\!\!\!\Uparrow^{_{\Varid{p}}}\!\!\;\con{I}_{\Varid{r}}\;{}\<[27]%
\>[27]{}\mathrel{=}\;\con{I}_{\Varid{p}}{}\<[E]%
\\
\>[3]{}{}_{\Varid{r}}\!\!\!\Uparrow^{_{\Varid{p}}}\!\!\;(\Conid{F}\mathbin{\consym{⊕}_{\Varid{r}}}\Conid{G})\;{}\<[27]%
\>[27]{}\mathrel{=}\;({}_{\Varid{r}}\!\!\!\Uparrow^{_{\Varid{p}}}\!\!\;\Conid{F})\mathbin{\consym{⊕}_{\Varid{p}}}({}_{\Varid{r}}\!\!\!\Uparrow^{_{\Varid{p}}}\!\!\;\Conid{G}){}\<[E]%
\\
\>[3]{}{}_{\Varid{r}}\!\!\!\Uparrow^{_{\Varid{p}}}\!\!\;(\Conid{F}\mathbin{\consym{⊗}_{\Varid{r}}}\Conid{G})\;{}\<[27]%
\>[27]{}\mathrel{=}\;({}_{\Varid{r}}\!\!\!\Uparrow^{_{\Varid{p}}}\!\!\;\Conid{F})\mathbin{\consym{⊗}_{\Varid{p}}}({}_{\Varid{r}}\!\!\!\Uparrow^{_{\Varid{p}}}\!\!\;\Conid{G}){}\<[E]%
\ColumnHook
\end{hscode}\resethooks
Since all libraries share similar names, we use subscripts to denote to
what library we are referring; \ensuremath{\Varid{r}} for \regular and \ensuremath{\Varid{p}} for \polyp. All
\regular codes embed trivially into \polyp, so \ensuremath{{}_{\Varid{r}}\!\!\!\Uparrow^{_{\Varid{p}}}\!\!} is unsurprising.
After having defined code conversion, we can show that the interpretation of
a code in \regular is equivalent to the converted code in \polyp. We do this
by defining an isomorphism between the two interpretations (and their fixed
points).
\Pedro{Define isomorphism.}%
We show one direction of the conversion, from \regular to \polyp:
\begin{hscode}\SaveRestoreHook
\column{B}{@{}>{\hspre}l<{\hspost}@{}}%
\column{3}{@{}>{\hspre}l<{\hspost}@{}}%
\column{35}{@{}>{\hspre}l<{\hspost}@{}}%
\column{42}{@{}>{\hspre}l<{\hspost}@{}}%
\column{48}{@{}>{\hspre}l<{\hspost}@{}}%
\column{56}{@{}>{\hspre}l<{\hspost}@{}}%
\column{72}{@{}>{\hspre}l<{\hspost}@{}}%
\column{E}{@{}>{\hspre}l<{\hspost}@{}}%
\>[3]{}\id{from}_{\Varid{r}}\;\mathbin{:}\;\{\mskip1.5mu \Conid{R}\;\mathbin{:}\;\ty{Set}\mskip1.5mu\}\;(\Conid{C}\;\mathbin{:}\;\ty{Code}_{\Varid{r}})\;\Varid{→}\;\id{⟦}\Conid{C}\id{⟧}_{\Varid{r}}\;\Conid{R}\;\Varid{→}\;\id{⟦}{}_{\Varid{r}}\!\!\!\Uparrow^{_{\Varid{p}}}\!\!\;\Conid{C}\id{⟧}_{\Varid{p}}\;\ty{⊥}\;\Conid{R}{}\<[E]%
\\
\>[3]{}\id{from}_{\Varid{r}}\;\con{U}_{\Varid{r}}\;{}\<[35]%
\>[35]{}\con{tt}\;{}\<[48]%
\>[48]{}\mathrel{=}\;\con{tt}{}\<[E]%
\\
\>[3]{}\id{from}_{\Varid{r}}\;\con{I}_{\Varid{r}}\;{}\<[35]%
\>[35]{}\Varid{x}\;{}\<[48]%
\>[48]{}\mathrel{=}\;\Varid{x}{}\<[E]%
\\
\>[3]{}\id{from}_{\Varid{r}}\;(\Conid{F}\mathbin{\consym{⊕}_{\Varid{r}}}\Conid{G})\;{}\<[35]%
\>[35]{}(\con{inj₁}\;{}\<[42]%
\>[42]{}\Varid{x})\;{}\<[48]%
\>[48]{}\mathrel{=}\;\con{inj₁}\;{}\<[56]%
\>[56]{}(\id{from}_{\Varid{r}}\;\Conid{F}\;{}\<[72]%
\>[72]{}\Varid{x}){}\<[E]%
\\
\>[3]{}\id{from}_{\Varid{r}}\;(\Conid{F}\mathbin{\consym{⊕}_{\Varid{r}}}\Conid{G})\;{}\<[35]%
\>[35]{}(\con{inj₂}\;{}\<[42]%
\>[42]{}\Varid{x})\;{}\<[48]%
\>[48]{}\mathrel{=}\;\con{inj₂}\;{}\<[56]%
\>[56]{}(\id{from}_{\Varid{r}}\;\Conid{G}\;{}\<[72]%
\>[72]{}\Varid{x}){}\<[E]%
\\
\>[3]{}\id{from}_{\Varid{r}}\;(\Conid{F}\mathbin{\consym{⊗}_{\Varid{r}}}\Conid{G})\;{}\<[35]%
\>[35]{}(\Varid{x}\;\consym{,}\,\;\Varid{y})\;{}\<[48]%
\>[48]{}\mathrel{=}\;\id{from}_{\Varid{r}}\;\Conid{F}\;\Varid{x}\;\consym{,}\,\;\id{from}_{\Varid{r}}\;\Conid{G}\;\Varid{y}{}\<[E]%
\\[\blanklineskip]%
\>[3]{}\Varid{fromμ}_{\Varid{r}}\;\mathbin{:}\;(\Conid{C}\;\mathbin{:}\;\ty{Code}_{\Varid{r}})\;\Varid{→}\;\ty{μ}_{\Varid{r}}\;\Conid{C}\;\Varid{→}\;\ty{μ}_{\Varid{p}}\;({}_{\Varid{r}}\!\!\!\Uparrow^{_{\Varid{p}}}\!\!\;\Conid{C})\;\ty{⊥}{}\<[E]%
\\
\>[3]{}\Varid{fromμ}_{\Varid{r}}\;\Conid{C}\;\con{⟨}\Varid{x}\con{⟩}_{\Varid{r}}\;\mathrel{=}\;\con{⟨}\id{from}_{\Varid{r}}\;\Conid{C}\;(\id{map}_{\Varid{r}}\;\Conid{C}\;(\Varid{fromμ}_{\Varid{r}}\;\Conid{C})\;\Varid{x})\con{⟩}_{\Varid{p}}{}\<[E]%
\ColumnHook
\end{hscode}\resethooks
Since \regular does not support parameters, we set the \polyp parameter to \ensuremath{\ty{⊥}}
for \regular codes. Function \ensuremath{\id{from}_{\Varid{r}}} does the conversion of one layer,
while \ensuremath{\Varid{fromμ}_{\Varid{r}}} ties the recursive knot by expanding fixed points,
converting one layer, and mapping itself recursively. Unfortunately
\ensuremath{\Varid{fromμ}_{\Varid{r}}} (and indeed all conversions in this paper involving fixed
points) does not pass Agda's termination checker; we provide some insights
on how to address this problem in \autoref{sec:discussion}.

The conversion in the other direction (\ensuremath{\id{to}_{\Varid{r}}} and \ensuremath{\Varid{toμ}_{\Varid{r}}})
is entirely symmetrical.

Having defined the conversion functions, we now have to prove that they indeed
form an isomorphism. We first consider the case without fixed points:
\begin{hscode}\SaveRestoreHook
\column{B}{@{}>{\hspre}l<{\hspost}@{}}%
\column{3}{@{}>{\hspre}l<{\hspost}@{}}%
\column{27}{@{}>{\hspre}l<{\hspost}@{}}%
\column{34}{@{}>{\hspre}l<{\hspost}@{}}%
\column{40}{@{}>{\hspre}l<{\hspost}@{}}%
\column{53}{@{}>{\hspre}l<{\hspost}@{}}%
\column{62}{@{}>{\hspre}l<{\hspost}@{}}%
\column{E}{@{}>{\hspre}l<{\hspost}@{}}%
\>[3]{}\id{iso₁}\;\mathbin{:}\;\{\mskip1.5mu \Conid{R}\;\mathbin{:}\;\ty{Set}\mskip1.5mu\}\;\Varid{→}\;(\Conid{C}\;\mathbin{:}\;\ty{Code}_{\Varid{r}})\;\Varid{→}\;(\Varid{x}\;\mathbin{:}\;\id{⟦}\Conid{C}\id{⟧}_{\Varid{r}}\;\Conid{R})\;\Varid{→}\;\id{to}_{\Varid{r}}\;\Conid{C}\;(\id{from}_{\Varid{r}}\;\Conid{C}\;\Varid{x})\;\ty{≡}\;\Varid{x}{}\<[E]%
\\
\>[3]{}\id{iso₁}\;\con{U}_{\Varid{r}}\;{}\<[27]%
\>[27]{}\__{}\;{}\<[40]%
\>[40]{}\mathrel{=}\;\con{refl}{}\<[E]%
\\
\>[3]{}\id{iso₁}\;\con{I}_{\Varid{r}}\;{}\<[27]%
\>[27]{}\__{}\;{}\<[40]%
\>[40]{}\mathrel{=}\;\con{refl}{}\<[E]%
\\
\>[3]{}\id{iso₁}\;(\Conid{F}\mathbin{\consym{⊕}_{\Varid{r}}}\Conid{G})\;{}\<[27]%
\>[27]{}(\con{inj₁}\;{}\<[34]%
\>[34]{}\Varid{x})\;{}\<[40]%
\>[40]{}\mathrel{=}\;\Varid{cong}\;\con{inj₁}\;{}\<[53]%
\>[53]{}(\id{iso₁}\;\Conid{F}\;{}\<[62]%
\>[62]{}\Varid{x}){}\<[E]%
\\
\>[3]{}\id{iso₁}\;(\Conid{F}\mathbin{\consym{⊕}_{\Varid{r}}}\Conid{G})\;{}\<[27]%
\>[27]{}(\con{inj₂}\;{}\<[34]%
\>[34]{}\Varid{x})\;{}\<[40]%
\>[40]{}\mathrel{=}\;\Varid{cong}\;\con{inj₂}\;{}\<[53]%
\>[53]{}(\id{iso₁}\;\Conid{G}\;{}\<[62]%
\>[62]{}\Varid{x}){}\<[E]%
\\
\>[3]{}\id{iso₁}\;(\Conid{F}\mathbin{\consym{⊗}_{\Varid{r}}}\Conid{G})\;{}\<[27]%
\>[27]{}(\Varid{x}\;\consym{,}\,\;\Varid{y})\;{}\<[40]%
\>[40]{}\mathrel{=}\;\Varid{cong₂}\;\__{}\consym{,}\,\__{}\;(\id{iso₁}\;\Conid{F}\;\Varid{x})\;(\id{iso₁}\;\Conid{G}\;\Varid{y}){}\<[E]%
\ColumnHook
\end{hscode}\resethooks
This proof is trivial, and so is its counterpart for the other direction
(\ensuremath{\id{from}_{\Varid{r}}\;\Conid{C}\;(\id{to}_{\Varid{r}}\;\Conid{C}\;\Varid{x})\;\ty{≡}\;\Varid{x}}). We use the Agda terms \ensuremath{\con{refl}}, \ensuremath{\Varid{sym}},
and \ensuremath{\Varid{cong}} for expressing reflexivity, symmetry, and congruence of equivalences,
respectively. In particular \ensuremath{\Varid{cong}} is used very often as it allows us to set
the focus of the proof deeper inside an expression.

When considering fixed points the proofs become more involved, since recursion
has to be taken into account. However, using the equational reasoning module
from the standard library (see the work of Shin-Cheng Mu et al.~\cite{AoPA}
for a detailed account on this style of proofs in Agda) we can keep the proofs
readable:

\begin{hscode}\SaveRestoreHook
\column{B}{@{}>{\hspre}l<{\hspost}@{}}%
\column{3}{@{}>{\hspre}l<{\hspost}@{}}%
\column{7}{@{}>{\hspre}l<{\hspost}@{}}%
\column{9}{@{}>{\hspre}l<{\hspost}@{}}%
\column{E}{@{}>{\hspre}l<{\hspost}@{}}%
\>[3]{}\Keyword{open}\;\Varid{≡-Reasoning}{}\<[E]%
\\[\blanklineskip]%
\>[3]{}\Varid{isoμ₁}\;\mathbin{:}\;(\Conid{C}\;\mathbin{:}\;\ty{Code}_{\Varid{r}})\;(\Varid{x}\;\mathbin{:}\;\ty{μ}_{\Varid{r}}\;\Conid{C})\;\Varid{→}\;\Varid{toμ}_{\Varid{r}}\;\Conid{C}\;(\Varid{fromμ}_{\Varid{r}}\;\Conid{C}\;\Varid{x})\;\ty{≡}\;\Varid{x}{}\<[E]%
\\
\>[3]{}\Varid{isoμ₁}\;\Conid{C}\;\con{⟨}\Varid{x}\con{⟩}_{\Varid{r}}\;\mathrel{=}\;\Varid{cong}\;\con{⟨}\_\con{⟩}_{\Varid{r}}\;\mathbin{\$}\;{}\<[E]%
\\[\blanklineskip]%
\>[3]{}\hsindent{4}{}\<[7]%
\>[7]{}\Varid{begin}\;{}\<[E]%
\\[\blanklineskip]%
\>[7]{}\hsindent{2}{}\<[9]%
\>[9]{}\highlight{\id{to}_{\Varid{r}}\;\Conid{C}\;(\id{map}_{\Varid{p}}\;({}_{\Varid{r}}\!\!\!\Uparrow^{_{\Varid{p}}}\!\!\;\Conid{C})\;\id{id}\;(\Varid{toμ}_{\Varid{r}}\;\Conid{C})\;}(\id{from}_{\Varid{r}}\;\Conid{C}\;(\id{map}_{\Varid{r}}\;\Conid{C}\;(\Varid{fromμ}_{\Varid{r}}\;\Conid{C})\;\Varid{x})))\;{}\<[E]%
\\[\blanklineskip]%
\>[3]{}\hsindent{4}{}\<[7]%
\>[7]{}\Varid{≡⟨}\;\Varid{mapCommute}_{\Varid{r}}^{_{\Varid{p}}}\;\Conid{C}\;\__{}\;\con{⟩}\;{}\<[E]%
\\[\blanklineskip]%
\>[7]{}\hsindent{2}{}\<[9]%
\>[9]{}\id{map}_{\Varid{r}}\;\Conid{C}\;(\Varid{toμ}_{\Varid{r}}\;\Conid{C})\;(\highlight{\id{to}_{\Varid{r}}\;\Conid{C}\;(\id{from}_{\Varid{r}}\;\Conid{C}\;}(\id{map}_{\Varid{r}}\;\Conid{C}\;(\Varid{fromμ}_{\Varid{r}}\;\Conid{C})\;\Varid{x})))\;{}\<[E]%
\\[\blanklineskip]%
\>[3]{}\hsindent{4}{}\<[7]%
\>[7]{}\Varid{≡⟨}\;\Varid{cong}\;(\id{map}_{\Varid{r}}\;\Conid{C}\;(\Varid{toμ}_{\Varid{r}}\;\Conid{C}))\;(\id{iso₁}\;\Conid{C}\;\__{})\;\con{⟩}\;{}\<[E]%
\\[\blanklineskip]%
\>[7]{}\hsindent{2}{}\<[9]%
\>[9]{}\highlight{\id{map}_{\Varid{r}}\;\Conid{C}\;}(\Varid{toμ}_{\Varid{r}}\;\Conid{C})\;(\highlight{\id{map}_{\Varid{r}}\;\Conid{C}\;}(\Varid{fromμ}_{\Varid{r}}\;\Conid{C})\;\Varid{x})\;{}\<[E]%
\\[\blanklineskip]%
\>[3]{}\hsindent{4}{}\<[7]%
\>[7]{}\Varid{≡⟨}\;\id{map}^{\mathbin{\circ}}_{\Varid{r}}\;\Conid{C}\;\con{⟩}\;{}\<[E]%
\\[\blanklineskip]%
\>[7]{}\hsindent{2}{}\<[9]%
\>[9]{}\id{map}_{\Varid{r}}\;\Conid{C}\;\highlight{(\Varid{toμ}_{\Varid{r}}\;\Conid{C}\;\id{∘}\;\Varid{fromμ}_{\Varid{r}}\;\Conid{C})\;}\Varid{x}\;{}\<[E]%
\\[\blanklineskip]%
\>[3]{}\hsindent{4}{}\<[7]%
\>[7]{}\Varid{≡⟨}\;\id{map}^{\Varid{∀}}_{\Varid{r}}\;\Conid{C}\;(\Varid{isoμ₁}\;\Conid{C})\;\Varid{x}\;\con{⟩}\;{}\<[E]%
\\[\blanklineskip]%
\>[7]{}\hsindent{2}{}\<[9]%
\>[9]{}\highlight{\id{map}_{\Varid{r}}\;\Conid{C}\;\id{id}\;}\Varid{x}\;{}\<[E]%
\\[\blanklineskip]%
\>[3]{}\hsindent{4}{}\<[7]%
\>[7]{}\Varid{≡⟨}\;\id{map}^{\id{id}}_{\Varid{r}}\;\Conid{C}\;\con{⟩}\;{}\<[E]%
\\[\blanklineskip]%
\>[7]{}\hsindent{2}{}\<[9]%
\>[9]{}\Varid{x}\;\qed{}\<[E]%
\ColumnHook
\end{hscode}\resethooks
To ease the reading of the equational style proofs, we highlight the term(s)
that we are focusing on at each step. In this proof we start with an argument
relating the \ensuremath{\id{map}}s of \regular and \polyp:
\begin{hscode}\SaveRestoreHook
\column{B}{@{}>{\hspre}l<{\hspost}@{}}%
\column{3}{@{}>{\hspre}l<{\hspost}@{}}%
\column{11}{@{}>{\hspre}l<{\hspost}@{}}%
\column{14}{@{}>{\hspre}l<{\hspost}@{}}%
\column{E}{@{}>{\hspre}l<{\hspost}@{}}%
\>[3]{}\Varid{mapCommute}_{\Varid{r}}^{_{\Varid{p}}}\;{}\<[11]%
\>[11]{}\mathbin{:}\;{}\<[14]%
\>[14]{}\{\mskip1.5mu \Conid{R₁}\;\Conid{R₂}\;\mathbin{:}\;\ty{Set}\mskip1.5mu\}\;\{\mskip1.5mu \Varid{f}\;\mathbin{:}\;\Conid{R₁}\;\Varid{→}\;\Conid{R₂}\mskip1.5mu\}\;(\Conid{C}\;\mathbin{:}\;\ty{Code}_{\Varid{r}})\;(\Varid{x}\;\mathbin{:}\;\id{⟦}{}_{\Varid{r}}\!\!\!\Uparrow^{_{\Varid{p}}}\!\!\;\Conid{C}\id{⟧}_{\Varid{p}}\;\ty{⊥}\;\Conid{R₁})\;{}\<[E]%
\\
\>[11]{}\Varid{→}\;{}\<[14]%
\>[14]{}\id{to}_{\Varid{r}}\;\Conid{C}\;(\id{map}_{\Varid{p}}\;({}_{\Varid{r}}\!\!\!\Uparrow^{_{\Varid{p}}}\!\!\;\Conid{C})\;\id{id}\;\Varid{f}\;\Varid{x})\;\ty{≡}\;\id{map}_{\Varid{r}}\;\Conid{C}\;\Varid{f}\;(\id{to}_{\Varid{r}}\;\Conid{C}\;\Varid{x}){}\<[E]%
\ColumnHook
\end{hscode}\resethooks
In words, this theorem states that the following two operations over a \regular
term \ensuremath{\Varid{x}} that has been lifted into \polyp are equivalent:
\begin{itemize}
\item To map a function \ensuremath{\Varid{f}} over the recursive positions of \ensuremath{\Varid{x}} in \polyp
      with \ensuremath{\id{map}_{\Varid{p}}} and then convert to \regular;
\item To first convert \ensuremath{\Varid{x}} back into \regular, and then map the function \ensuremath{\Varid{f}}
      with \ensuremath{\id{map}_{\Varid{r}}}.
\end{itemize}
After this step, we either proceed by a recursive
argument (referring to \ensuremath{\id{iso₁}} or \ensuremath{\Varid{isoμ₁}}) or by a lemma. For conciseness, we
show only the types of the lemmas:
\begin{hscode}\SaveRestoreHook
\column{B}{@{}>{\hspre}l<{\hspost}@{}}%
\column{3}{@{}>{\hspre}l<{\hspost}@{}}%
\column{17}{@{}>{\hspre}l<{\hspost}@{}}%
\column{20}{@{}>{\hspre}l<{\hspost}@{}}%
\column{E}{@{}>{\hspre}l<{\hspost}@{}}%
\>[3]{}\id{map}^{\mathbin{\circ}}_{\Varid{r}}\;{}\<[17]%
\>[17]{}\mathbin{:}\;{}\<[20]%
\>[20]{}\{\mskip1.5mu \Conid{A}\;\Conid{B}\;\Conid{C}\;\mathbin{:}\;\ty{Set}\mskip1.5mu\}\;\{\mskip1.5mu \Varid{f}\;\mathbin{:}\;\Conid{B}\;\Varid{→}\;\Conid{C}\mskip1.5mu\}\;\{\mskip1.5mu \Varid{g}\;\mathbin{:}\;\Conid{A}\;\Varid{→}\;\Conid{B}\mskip1.5mu\}\;(\Conid{D}\;\mathbin{:}\;\ty{Code}_{\Varid{r}})\;\{\mskip1.5mu \Varid{x}\;\mathbin{:}\;\id{⟦}\Conid{D}\id{⟧}_{\Varid{r}}\;\Conid{A}\mskip1.5mu\}\;{}\<[E]%
\\
\>[17]{}\Varid{→}\;{}\<[20]%
\>[20]{}\id{map}_{\Varid{r}}\;\Conid{D}\;\Varid{f}\;(\id{map}_{\Varid{r}}\;\Conid{D}\;\Varid{g}\;\Varid{x})\;\ty{≡}\;\id{map}_{\Varid{r}}\;\Conid{D}\;(\Varid{f}\;\id{∘}\;\Varid{g})\;\Varid{x}{}\<[E]%
\\[\blanklineskip]%
\>[3]{}\id{map}^{\Varid{∀}}_{\Varid{r}}\;{}\<[17]%
\>[17]{}\mathbin{:}\;{}\<[20]%
\>[20]{}(\Conid{C}\;\mathbin{:}\;\ty{Code}_{\Varid{r}})\;\{\mskip1.5mu \Conid{A}\;\Conid{B}\;\mathbin{:}\;\ty{Set}\mskip1.5mu\}\;\{\mskip1.5mu \Varid{f}\;\Varid{g}\;\mathbin{:}\;\Conid{A}\;\Varid{→}\;\Conid{B}\mskip1.5mu\}\;{}\<[E]%
\\
\>[17]{}\Varid{→}\;{}\<[20]%
\>[20]{}(\Varid{∀}\;\Varid{x}\;\Varid{→}\;\Varid{f}\;\Varid{x}\;\ty{≡}\;\Varid{g}\;\Varid{x})\;\Varid{→}\;(\Varid{∀}\;\Varid{x}\;\Varid{→}\;\id{map}_{\Varid{r}}\;\Conid{C}\;\Varid{f}\;\Varid{x}\;\ty{≡}\;\id{map}_{\Varid{r}}\;\Conid{C}\;\Varid{g}\;\Varid{x}){}\<[E]%
\\[\blanklineskip]%
\>[3]{}\id{map}^{\id{id}}_{\Varid{r}}\;\mathbin{:}\;\Varid{∀}\;\{\mskip1.5mu \Conid{A}\mskip1.5mu\}\;(\Conid{C}\;\mathbin{:}\;\ty{Code}_{\Varid{r}})\;\{\mskip1.5mu \Varid{x}\;\mathbin{:}\;\id{⟦}\Conid{C}\id{⟧}_{\Varid{r}}\;\Conid{A}\mskip1.5mu\}\;\Varid{→}\;\id{map}_{\Varid{r}}\;\Conid{C}\;\id{id}\;\Varid{x}\;\ty{≡}\;\Varid{x}{}\<[E]%
\ColumnHook
\end{hscode}\resethooks
These lemmas are standard properties of \ensuremath{\id{map}_{\Varid{r}}}, namely the functor laws
and the fact that \ensuremath{\id{map}_{\Varid{r}}} preserves extensional equality (a form of
congruence on \ensuremath{\id{map}_{\Varid{r}}}).
All of them are easily proved by induction on the codes.

Put together, \ensuremath{\Varid{fromμ}_{\Varid{r}}}, \ensuremath{\Varid{toμ}_{\Varid{r}}}, \ensuremath{\Varid{isoμ₁}}, and \ensuremath{\Varid{isoμ₂}} (the dual of
\ensuremath{\Varid{isoμ₁}}) form an isomorphism that shows how to embed \regular codes into
\polyp codes.

\subsection{PolyP to Indexed}
\label{sec:PolyP2Indexed}%

We proceed to the conversion between \polyp and \indexed codes. As we
mentioned before, particular care has to be taken with composition; the
remaining codes are trivially converted, so we only show composition:

\begin{hscode}\SaveRestoreHook
\column{B}{@{}>{\hspre}l<{\hspost}@{}}%
\column{3}{@{}>{\hspre}l<{\hspost}@{}}%
\column{22}{@{}>{\hspre}l<{\hspost}@{}}%
\column{28}{@{}>{\hspre}l<{\hspost}@{}}%
\column{E}{@{}>{\hspre}l<{\hspost}@{}}%
\>[3]{}{}_{\Varid{p}}\!\!\Uparrow^{_{\kern0.05em \Varid{i}}}\!\!\;\mathbin{:}\;\ty{Code}_{\Varid{p}}\;\Varid{→}\;\ty{Code}_{\kern0.05em \Varid{i}}\;(\ty{⊤}\;\ty{⊎}\;\ty{⊤})\;\ty{⊤}{}\<[E]%
\\[\blanklineskip]%
\>[3]{}{}_{\Varid{p}}\!\!\Uparrow^{_{\kern0.05em \Varid{i}}}\!\!\;(\Conid{F}\mathbin{\consym{⊚}_{\Varid{p}}}{}\<[22]%
\>[22]{}\Conid{G})\;{}\<[28]%
\>[28]{}\mathrel{=}\;(\con{Fix}_{\kern0.05em \Varid{i}}\;({}_{\Varid{p}}\!\!\Uparrow^{_{\kern0.05em \Varid{i}}}\!\!\;\Conid{F}))\mathbin{\consym{⊚}_{\kern0.05em \Varid{i}}}({}_{\Varid{p}}\!\!\Uparrow^{_{\kern0.05em \Varid{i}}}\!\!\;\Conid{G}){}\<[E]%
\ColumnHook
\end{hscode}\resethooks
We cannot simply take the \indexed composition of the two converted codes
because their types do not allow composition. A \polyp code results in an open
\indexed code with one parameter and one recursive position, therefore of type
\ensuremath{\ty{Code}_{\kern0.05em \Varid{i}}\;(\ty{⊤}\;\ty{⊎}\;\ty{⊤})\;\ty{⊤}}. Taking the fixed point of such a code gives a code
of type \ensuremath{\ty{Code}_{\kern0.05em \Varid{i}}\;\ty{⊤}\;\ty{⊤}}, so we can mimic \polyp's interpretation of
composition in our conversion, using the \ensuremath{\con{Fix}_{\kern0.05em \Varid{i}}} of \indexed.
In fact, converting \polyp to \indexed helps us understand the interpretation
of composition in \polyp, because the types now show us that there's no way
to define a composition other than by combining it with the fixed-point 
operator.

Converting composed values from \polyp is then a recursive task, due to the
presence of fixed points. We first convert the outer functor \ensuremath{\Conid{F}}, and then
map the conversion onto the arguments, recalling that on the left we have
parameter codes \ensuremath{\Conid{G}}, while on the right we have recursive occurrences of the
original composition:

\begin{hscode}\SaveRestoreHook
\column{B}{@{}>{\hspre}l<{\hspost}@{}}%
\column{3}{@{}>{\hspre}l<{\hspost}@{}}%
\column{28}{@{}>{\hspre}l<{\hspost}@{}}%
\column{43}{@{}>{\hspre}l<{\hspost}@{}}%
\column{46}{@{}>{\hspre}l<{\hspost}@{}}%
\column{E}{@{}>{\hspre}l<{\hspost}@{}}%
\>[3]{}\id{from}_{\Varid{p}}\;\mathbin{:}\;\{\mskip1.5mu \Conid{A}\;\Conid{R}\;\mathbin{:}\;\ty{Set}\mskip1.5mu\}\;(\Conid{C}\;\mathbin{:}\;\ty{Code}_{\Varid{p}})\;\Varid{→}\;\id{⟦}\Conid{C}\id{⟧}_{\Varid{p}}\;\Conid{A}\;\Conid{R}\;\Varid{→}\;\id{⟦}{}_{\Varid{p}}\!\!\Uparrow^{_{\kern0.05em \Varid{i}}}\!\!\;\Conid{C}\id{⟧}_{\kern0.05em \Varid{i}}\;((\id{const}\;\Conid{A})\mathbin{\idsym{∣}_{\kern0.05em \Varid{i}}}(\id{const}\;\Conid{R}))\;\con{tt}{}\<[E]%
\\[\blanklineskip]%
\>[3]{}\id{from}_{\Varid{p}}\;(\Conid{F}\mathbin{\consym{⊚}_{\Varid{p}}}\Conid{G})\;{}\<[28]%
\>[28]{}\con{⟨}\Varid{x}\con{⟩}_{\Varid{p}}\;{}\<[43]%
\>[43]{}\mathrel{=}\;{}\<[46]%
\>[46]{}\con{⟨}\id{map}_{\kern0.05em \Varid{i}}\;({}_{\Varid{p}}\!\!\Uparrow^{_{\kern0.05em \Varid{i}}}\!\!\;\Conid{F})\;((\Varid{λ}\;\__{}\;\Varid{→}\;\id{from}_{\Varid{p}}\;\Conid{G})\mathbin{\idsym{∥}_{\kern0.05em \Varid{i}}}(\Varid{λ}\;\__{}\;\Varid{→}\;\id{from}_{\Varid{p}}\;(\Conid{F}\mathbin{\consym{⊚}_{\Varid{p}}}\Conid{G})))\;\con{tt}\;(\id{from}_{\Varid{p}}\;\Conid{F}\;\Varid{x})\con{⟩}_{\kern0.05em \Varid{i}}{}\<[E]%
\ColumnHook
\end{hscode}\resethooks

We also show the conversion in the opposite direction, which is entirely
symmetrical:

\begin{hscode}\SaveRestoreHook
\column{B}{@{}>{\hspre}l<{\hspost}@{}}%
\column{3}{@{}>{\hspre}l<{\hspost}@{}}%
\column{26}{@{}>{\hspre}l<{\hspost}@{}}%
\column{43}{@{}>{\hspre}l<{\hspost}@{}}%
\column{E}{@{}>{\hspre}l<{\hspost}@{}}%
\>[3]{}\id{to}_{\Varid{p}}\;\mathbin{:}\;\{\mskip1.5mu \Conid{A}\;\Conid{R}\;\mathbin{:}\;\ty{Set}\mskip1.5mu\}\;(\Conid{C}\;\mathbin{:}\;\ty{Code}_{\Varid{p}})\;\Varid{→}\;\id{⟦}{}_{\Varid{p}}\!\!\Uparrow^{_{\kern0.05em \Varid{i}}}\!\!\;\Conid{C}\id{⟧}_{\kern0.05em \Varid{i}}\;((\id{const}\;\Conid{A})\mathbin{\idsym{∣}_{\kern0.05em \Varid{i}}}(\id{const}\;\Conid{R}))\;\con{tt}\;\Varid{→}\;\id{⟦}\Conid{C}\id{⟧}_{\Varid{p}}\;\Conid{A}\;\Conid{R}{}\<[E]%
\\[\blanklineskip]%
\>[3]{}\id{to}_{\Varid{p}}\;(\Conid{F}\mathbin{\consym{⊚}_{\Varid{p}}}\Conid{G})\;{}\<[26]%
\>[26]{}\con{⟨}\Varid{x}\con{⟩}_{\kern0.05em \Varid{i}}\;{}\<[43]%
\>[43]{}\mathrel{=}\;\con{⟨}\id{to}_{\Varid{p}}\;\Conid{F}\;(\id{map}_{\kern0.05em \Varid{i}}\;({}_{\Varid{p}}\!\!\Uparrow^{_{\kern0.05em \Varid{i}}}\!\!\;\Conid{F})\;((\Varid{λ}\;\__{}\;\Varid{→}\;\id{to}_{\Varid{p}}\;\Conid{G})\mathbin{\idsym{∥}_{\kern0.05em \Varid{i}}}(\Varid{λ}\;\__{}\;\Varid{→}\;\id{to}_{\Varid{p}}\;(\Conid{F}\mathbin{\consym{⊚}_{\Varid{p}}}\Conid{G})))\;\con{tt}\;\Varid{x})\con{⟩}_{\Varid{p}}{}\<[E]%
\ColumnHook
\end{hscode}\resethooks

The conversion of \polyp fixed points is very similar to the conversion of
composition. The main difference lies in the functions that we map to the
arguments and recursive positions:

\begin{hscode}\SaveRestoreHook
\column{B}{@{}>{\hspre}l<{\hspost}@{}}%
\column{3}{@{}>{\hspre}l<{\hspost}@{}}%
\column{E}{@{}>{\hspre}l<{\hspost}@{}}%
\>[3]{}\Varid{fromμ}_{\Varid{p}}\;\mathbin{:}\;\{\mskip1.5mu \Conid{A}\;\mathbin{:}\;\ty{Set}\mskip1.5mu\}\;(\Conid{C}\;\mathbin{:}\;\ty{Code}_{\Varid{p}})\;\Varid{→}\;\ty{μ}_{\Varid{p}}\;\Conid{C}\;\Conid{A}\;\Varid{→}\;\id{⟦}\con{Fix}_{\kern0.05em \Varid{i}}\;({}_{\Varid{p}}\!\!\Uparrow^{_{\kern0.05em \Varid{i}}}\!\!\;\Conid{C})\id{⟧}_{\kern0.05em \Varid{i}}\;(\id{const}\;\Conid{A})\;\con{tt}{}\<[E]%
\\
\>[3]{}\Varid{fromμ}_{\Varid{p}}\;\Conid{C}\;\con{⟨}\Varid{x}\con{⟩}_{\Varid{p}}\;\mathrel{=}\;\con{⟨}\id{map}_{\kern0.05em \Varid{i}}\;({}_{\Varid{p}}\!\!\Uparrow^{_{\kern0.05em \Varid{i}}}\!\!\;\Conid{C})\;((\id{const}\;\id{id})\mathbin{\idsym{∥}_{\kern0.05em \Varid{i}}}(\Varid{λ}\;\__{}\;\Varid{→}\;\Varid{fromμ}_{\Varid{p}}\;\Conid{C}))\;\con{tt}\;(\id{from}_{\Varid{p}}\;\Conid{C}\;\Varid{x})\con{⟩}_{\kern0.05em \Varid{i}}{}\<[E]%
\ColumnHook
\end{hscode}\resethooks

We omit the conversion in the opposite direction for brevity, and also the
isomorphism proof. Both the case for composition and for \polyp fixed points
require lengthy proofs using properties of \ensuremath{\id{map}_{\kern0.05em \Varid{i}}}, which are presented
in more detail in the following section.%
\footnote{This proof and other omitted details can be found in the code
available at the first author's webpage (\url{http://dreixel.net}).}

\subsection{Indexed to InstantGenerics}
\label{sec:Indexed2IG}%

As a final example we show how to convert from a fixed-point view to the
coinductive representation of \ig. Since all fixed-point views embed into
\indexed, we need to define only the embedding of \indexed into \ig. Since
the two universes are less similar, the code transformation requires more care:

\begin{hscode}\SaveRestoreHook
\column{B}{@{}>{\hspre}l<{\hspost}@{}}%
\column{3}{@{}>{\hspre}l<{\hspost}@{}}%
\column{29}{@{}>{\hspre}l<{\hspost}@{}}%
\column{E}{@{}>{\hspre}l<{\hspost}@{}}%
\>[3]{}{}_{\kern0.05em \Varid{i}}\!\!\Uparrow^{_{\Varid{ig}}}\!\!\;\mathbin{:}\;\{\mskip1.5mu \Conid{I}\;\Conid{O}\;\mathbin{:}\;\ty{Set}\mskip1.5mu\}\;\Varid{→}\;\ty{Code}_{\kern0.05em \Varid{i}}\;\Conid{I}\;\Conid{O}\;\Varid{→}\;(\Conid{I}\;\Varid{→}\;\ty{Set})\;\Varid{→}\;(\Conid{O}\;\Varid{→}\;\ty{Code}_{\Varid{ig}}){}\<[E]%
\\
\>[3]{}{}_{\kern0.05em \Varid{i}}\!\!\Uparrow^{_{\Varid{ig}}}\!\!\;\con{U}_{\kern0.05em \Varid{i}}\;{}\<[29]%
\>[29]{}\Varid{r}\;\Varid{o}\;\mathrel{=}\;\con{U}_{\Varid{ig}}{}\<[E]%
\\
\>[3]{}{}_{\kern0.05em \Varid{i}}\!\!\Uparrow^{_{\Varid{ig}}}\!\!\;(\con{I}_{\kern0.05em \Varid{i}}\;\Varid{i})\;{}\<[29]%
\>[29]{}\Varid{r}\;\Varid{o}\;\mathrel{=}\;\con{K}_{\Varid{ig}}\;(\Varid{r}\;\Varid{i}){}\<[E]%
\\
\>[3]{}{}_{\kern0.05em \Varid{i}}\!\!\Uparrow^{_{\Varid{ig}}}\!\!\;(\con{!}_{\kern0.05em \Varid{i}}\;\Varid{i})\;{}\<[29]%
\>[29]{}\Varid{r}\;\Varid{o}\;\mathrel{=}\;\con{K}_{\Varid{ig}}\;(\Varid{o}\;\ty{≡}\;\Varid{i}){}\<[E]%
\\[\blanklineskip]%
\>[3]{}{}_{\kern0.05em \Varid{i}}\!\!\Uparrow^{_{\Varid{ig}}}\!\!\;(\Conid{F}\mathbin{\consym{⊕}_{\kern0.05em \Varid{i}}}\Conid{G})\;{}\<[29]%
\>[29]{}\Varid{r}\;\Varid{o}\;\mathrel{=}\;({}_{\kern0.05em \Varid{i}}\!\!\Uparrow^{_{\Varid{ig}}}\!\!\;\Conid{F}\;\Varid{r}\;\Varid{o})\mathbin{\consym{⊕}_{\Varid{ig}}}({}_{\kern0.05em \Varid{i}}\!\!\Uparrow^{_{\Varid{ig}}}\!\!\;\Conid{G}\;\Varid{r}\;\Varid{o}){}\<[E]%
\\
\>[3]{}{}_{\kern0.05em \Varid{i}}\!\!\Uparrow^{_{\Varid{ig}}}\!\!\;(\Conid{F}\mathbin{\consym{⊗}_{\kern0.05em \Varid{i}}}\Conid{G})\;{}\<[29]%
\>[29]{}\Varid{r}\;\Varid{o}\;\mathrel{=}\;({}_{\kern0.05em \Varid{i}}\!\!\Uparrow^{_{\Varid{ig}}}\!\!\;\Conid{F}\;\Varid{r}\;\Varid{o})\mathbin{\consym{⊗}_{\Varid{ig}}}({}_{\kern0.05em \Varid{i}}\!\!\Uparrow^{_{\Varid{ig}}}\!\!\;\Conid{G}\;\Varid{r}\;\Varid{o}){}\<[E]%
\\
\>[3]{}{}_{\kern0.05em \Varid{i}}\!\!\Uparrow^{_{\Varid{ig}}}\!\!\;(\Conid{F}\mathbin{\consym{⊚}_{\kern0.05em \Varid{i}}}\Conid{G})\;{}\<[29]%
\>[29]{}\Varid{r}\;\Varid{o}\;\mathrel{=}\;\con{R}_{\Varid{ig}}\;(\Varid{♯}\;{}_{\kern0.05em \Varid{i}}\!\!\Uparrow^{_{\Varid{ig}}}\!\!\;\Conid{F}\;(\Varid{λ}\;\Varid{i}\;\Varid{→}\;\id{⟦}{}_{\kern0.05em \Varid{i}}\!\!\Uparrow^{_{\Varid{ig}}}\!\!\;\Conid{G}\;\Varid{r}\;\Varid{i}\id{⟧}_{\Varid{ig}})\;\Varid{o}){}\<[E]%
\\[\blanklineskip]%
\>[3]{}{}_{\kern0.05em \Varid{i}}\!\!\Uparrow^{_{\Varid{ig}}}\!\!\;(\con{Fix}_{\kern0.05em \Varid{i}}\;\Conid{F})\;{}\<[29]%
\>[29]{}\Varid{r}\;\Varid{o}\;\mathrel{=}\;\con{R}_{\Varid{ig}}\;(\Varid{♯}\;{}_{\kern0.05em \Varid{i}}\!\!\Uparrow^{_{\Varid{ig}}}\!\!\;\Conid{F}\;(\Varid{r}\mathbin{\idsym{∣}_{\kern0.05em \Varid{i}}}(\Varid{λ}\;\Varid{i}\;\Varid{→}\;\id{⟦}{}_{\kern0.05em \Varid{i}}\!\!\Uparrow^{_{\Varid{ig}}}\!\!\;(\con{Fix}_{\kern0.05em \Varid{i}}\;\Conid{F})\;\Varid{r}\;\Varid{i}\id{⟧}_{\Varid{ig}}))\;\Varid{o}){}\<[E]%
\ColumnHook
\end{hscode}\resethooks
Unit, sum, and product exist in both universes, so their conversion is
trivial. Recursive invocations with \ensuremath{\con{I}_{\kern0.05em \Varid{i}}} are replaced by simple
constants; we lose the ability to abstract over recursive positions, which is
in line with the behavior of \ig.
Tagging is also converted to a constant, trivially inhabited if we are in the
expected output index \ensuremath{\Varid{o}}, and empty otherwise. Note that since an \indexed code
can define multiple types, but an \ig code can only represent one type,
\ensuremath{{}_{\kern0.05em \Varid{i}}\!\!\Uparrow^{_{\Varid{ig}}}\!\!} \xspace effectively produces multiple \ig codes, one for each output
index of the original \indexed family.

A composition \ensuremath{\Conid{F}\mathbin{\consym{⊚}_{\kern0.05em \Varid{i}}}\Conid{G}} is encoded through recursion; the resulting
code is the conversion of \ensuremath{\Conid{F}}, whose parameters are \indexed \ensuremath{\Conid{G}} functors.
We convert these functors to sets using \ensuremath{{}_{\kern0.05em \Varid{i}}\!\!\Uparrow^{_{\Varid{ig}}}\!\!} \xspace to get \ig codes, which
we then interpret with \ensuremath{\id{⟦}\_\id{⟧}_{\Varid{ig}}}.

A fixed-point \ensuremath{\con{Fix}_{\kern0.05em \Varid{i}}\;\Conid{F}} is naturally encoded through recursion, in a
similar way to composition. The recursive positions of the fixed point are
either: parameters on the left, converted with \ensuremath{\Varid{r}} as before; or recursive
occurrences on the right, handled by recursively converting the codes with
\ensuremath{{}_{\kern0.05em \Varid{i}}\!\!\Uparrow^{_{\Varid{ig}}}\!\!} \xspace and interpreting.

Note that both for composition and fixed points we instantiate the function
which interprets indices (the \ensuremath{\Varid{r}} argument) with an \ig interpretation.
However, \ensuremath{\Varid{r}} has type \ensuremath{\Conid{I}\;\Varid{→}\;\ty{Set}}, whereas \ensuremath{\id{⟦}\_\id{⟧}_{\Varid{ig}}} has return type \ensuremath{\ty{Set₁}}.
If we were to raise \indexed to \ensuremath{\ty{Set₁}}, the
interpretation function would then have type \ensuremath{\Conid{I}\;\Varid{→}\;\ty{Set₁}}, but then we could
no longer use it in the \ensuremath{\con{I}_{\kern0.05em \Varid{i}}} case. For now we rely on the Agda flag
\text{\tt \char45{}\char45{}type\char45{}in\char45{}type}, and leave a formally correct solution for future work
(see \autoref{sec:discussion}).

Having the code conversion in place, we can proceed to convert values:
\begin{hscode}\SaveRestoreHook
\column{B}{@{}>{\hspre}l<{\hspost}@{}}%
\column{3}{@{}>{\hspre}l<{\hspost}@{}}%
\column{29}{@{}>{\hspre}l<{\hspost}@{}}%
\column{38}{@{}>{\hspre}l<{\hspost}@{}}%
\column{48}{@{}>{\hspre}l<{\hspost}@{}}%
\column{59}{@{}>{\hspre}l<{\hspost}@{}}%
\column{68}{@{}>{\hspre}l<{\hspost}@{}}%
\column{E}{@{}>{\hspre}l<{\hspost}@{}}%
\>[3]{}\id{from}\;\mathbin{:}\;\{\mskip1.5mu \Conid{I}\;\Conid{O}\;\mathbin{:}\;\ty{Set}\mskip1.5mu\}\;\{\mskip1.5mu \Varid{r}\;\mathbin{:}\;\Conid{I}\;\Varid{→}\;\ty{Set}\mskip1.5mu\}\;(\Conid{C}\;\mathbin{:}\;\ty{Code}_{\kern0.05em \Varid{i}}\;\Conid{I}\;\Conid{O})\;(\Varid{o}\;\mathbin{:}\;\Conid{O})\;\Varid{→}\;\id{⟦}\Conid{C}\id{⟧}_{\kern0.05em \Varid{i}}\;\Varid{r}\;\Varid{o}\;\Varid{→}\;\id{⟦}{}_{\kern0.05em \Varid{i}}\!\!\Uparrow^{_{\Varid{ig}}}\!\!\;\Conid{C}\;\Varid{r}\;\Varid{o}\id{⟧}_{\Varid{ig}}{}\<[E]%
\\[\blanklineskip]%
\>[3]{}\id{from}\;\con{U}_{\kern0.05em \Varid{i}}\;{}\<[29]%
\>[29]{}\Varid{o}\;\con{tt}\;{}\<[48]%
\>[48]{}\mathrel{=}\;\con{tt}_{\Varid{ig}}{}\<[E]%
\\
\>[3]{}\id{from}\;(\con{I}_{\kern0.05em \Varid{i}}\;\Varid{i})\;{}\<[29]%
\>[29]{}\Varid{o}\;\Varid{x}\;{}\<[48]%
\>[48]{}\mathrel{=}\;\con{k}_{\Varid{ig}}\;\Varid{x}{}\<[E]%
\\
\>[3]{}\id{from}\;(\con{!}_{\kern0.05em \Varid{i}}\;\Varid{i})\;{}\<[29]%
\>[29]{}\Varid{o}\;\Varid{x}\;{}\<[48]%
\>[48]{}\mathrel{=}\;\con{k}_{\Varid{ig}}\;\Varid{x}{}\<[E]%
\\[\blanklineskip]%
\>[3]{}\id{from}\;(\Conid{F}\mathbin{\consym{⊕}_{\kern0.05em \Varid{i}}}\Conid{G})\;{}\<[29]%
\>[29]{}\Varid{o}\;(\con{inj₁}\;{}\<[38]%
\>[38]{}\Varid{x})\;{}\<[48]%
\>[48]{}\mathrel{=}\;\con{inj₁}_{\Varid{ig}}\;{}\<[59]%
\>[59]{}(\id{from}\;\Conid{F}\;{}\<[68]%
\>[68]{}\Varid{o}\;\Varid{x}){}\<[E]%
\\
\>[3]{}\id{from}\;(\Conid{F}\mathbin{\consym{⊕}_{\kern0.05em \Varid{i}}}\Conid{G})\;{}\<[29]%
\>[29]{}\Varid{o}\;(\con{inj₂}\;{}\<[38]%
\>[38]{}\Varid{x})\;{}\<[48]%
\>[48]{}\mathrel{=}\;\con{inj₂}_{\Varid{ig}}\;{}\<[59]%
\>[59]{}(\id{from}\;\Conid{G}\;{}\<[68]%
\>[68]{}\Varid{o}\;\Varid{x}){}\<[E]%
\\
\>[3]{}\id{from}\;(\Conid{F}\mathbin{\consym{⊗}_{\kern0.05em \Varid{i}}}\Conid{G})\;{}\<[29]%
\>[29]{}\Varid{o}\;(\Varid{x}\;\consym{,}\,\;\Varid{y})\;{}\<[48]%
\>[48]{}\mathrel{=}\;(\id{from}\;\Conid{F}\;\Varid{o}\;\Varid{x}),_{\Varid{ig}}(\id{from}\;\Conid{G}\;\Varid{o}\;\Varid{y}){}\<[E]%
\\
\>[3]{}\id{from}\;(\Conid{F}\mathbin{\consym{⊚}_{\kern0.05em \Varid{i}}}\Conid{G})\;{}\<[29]%
\>[29]{}\Varid{o}\;\Varid{x}\;{}\<[48]%
\>[48]{}\mathrel{=}\;\con{rec}_{\Varid{ig}}\;(\id{from}\;\Conid{F}\;\Varid{o}\;(\id{map}_{\kern0.05em \Varid{i}}\;\Conid{F}\;(\id{from}\;\Conid{G})\;\Varid{o}\;\Varid{x})){}\<[E]%
\\[\blanklineskip]%
\>[3]{}\id{from}\;(\con{Fix}_{\kern0.05em \Varid{i}}\;\Conid{F})\;{}\<[29]%
\>[29]{}\Varid{o}\;\con{⟨}\Varid{x}\con{⟩}_{\kern0.05em \Varid{i}}\;{}\<[48]%
\>[48]{}\mathrel{=}\;\con{rec}_{\Varid{ig}}\;(\id{from}\;\Conid{F}\;\Varid{o}\;(\id{map}_{\kern0.05em \Varid{i}}\;\Conid{F}\;((\Varid{λ}\;\__{}\;\Varid{→}\;\id{id})\mathbin{\idsym{∥}_{\kern0.05em \Varid{i}}}(\id{from}\;(\con{Fix}_{\kern0.05em \Varid{i}}\;\Conid{F})))\;\Varid{o}\;\Varid{x})){}\<[E]%
\ColumnHook
\end{hscode}\resethooks
The cases for composition and fixed point are more challenging because we have
to map the conversion function inside the argument positions; we do this using
the \ensuremath{\id{map}_{\kern0.05em \Varid{i}}} function. As usual, the inverse function \ensuremath{\id{to}} is entirely
symmetrical, so we omit it.

It remains to show that the conversion functions form an isomorphism. We show
the only two interesting cases: composition and fixed points. Following
previous work~\cite{jpm:gpif:11}, we lift composition, equality, and identity
to natural transformations in \indexed (respectively \ensuremath{\_\mathbin{\idsym{∘\ensuremath{_⇉}}_{\kern0.05em \Varid{i}}}\_},
\ensuremath{\_\mathbin{\idsym{≅}_{\kern0.05em \Varid{i}}}\_}, and \ensuremath{\id{id\ensuremath{_⇉}}_{\kern0.05em \Varid{i}}}). We use equational
reasoning for the proofs, and highlight the part of the term that we
focus on in each line of the proof:

\begin{hscode}\SaveRestoreHook
\column{B}{@{}>{\hspre}l<{\hspost}@{}}%
\column{3}{@{}>{\hspre}l<{\hspost}@{}}%
\column{5}{@{}>{\hspre}l<{\hspost}@{}}%
\column{7}{@{}>{\hspre}l<{\hspost}@{}}%
\column{8}{@{}>{\hspre}l<{\hspost}@{}}%
\column{29}{@{}>{\hspre}l<{\hspost}@{}}%
\column{43}{@{}>{\hspre}l<{\hspost}@{}}%
\column{E}{@{}>{\hspre}l<{\hspost}@{}}%
\>[3]{}\id{iso₁}\;\mathbin{:}\;\{\mskip1.5mu \Conid{I}\;\Conid{O}\;\mathbin{:}\;\ty{Set}\mskip1.5mu\}\;(\Conid{C}\;\mathbin{:}\;\ty{Code}_{\kern0.05em \Varid{i}}\;\Conid{I}\;\Conid{O})\;(\Varid{r}\;\mathbin{:}\;\Conid{I}\;\Varid{→}\;\ty{Set})\;\Varid{→}\;(\id{to}\;\{\mskip1.5mu \Varid{r}\;\mathrel{=}\;\Varid{r}\mskip1.5mu\}\;\Conid{C}\mathbin{\idsym{∘\ensuremath{_⇉}}_{\kern0.05em \Varid{i}}}\id{from}\;\Conid{C})\mathbin{\idsym{≅}_{\kern0.05em \Varid{i}}}\id{id\ensuremath{_⇉}}_{\kern0.05em \Varid{i}}{}\<[E]%
\\[\blanklineskip]%
\>[3]{}\id{iso₁}\;(\Conid{F}\mathbin{\consym{⊚}_{\kern0.05em \Varid{i}}}\Conid{G})\;{}\<[29]%
\>[29]{}\Varid{r}\;\Varid{o}\;\Varid{x}\;{}\<[43]%
\>[43]{}\mathrel{=}\;{}\<[E]%
\\[\blanklineskip]%
\>[3]{}\hsindent{2}{}\<[5]%
\>[5]{}\Varid{begin}\;{}\<[E]%
\\[\blanklineskip]%
\>[5]{}\hsindent{2}{}\<[7]%
\>[7]{}\id{map}_{\kern0.05em \Varid{i}}\;\Conid{F}\;(\id{to}\;\Conid{G})\;\Varid{o}\;(\highlight{\id{to}\;\Conid{F}\;\Varid{o}\;(\id{from}\;\Conid{F}\;}\Varid{o}\;(\id{map}_{\kern0.05em \Varid{i}}\;\Conid{F}\;(\id{from}\;\Conid{G})\;\Varid{o}\;\Varid{x})))\;{}\<[E]%
\\[\blanklineskip]%
\>[3]{}\hsindent{2}{}\<[5]%
\>[5]{}\Varid{≡⟨}\;\Varid{cong}\;(\id{map}_{\kern0.05em \Varid{i}}\;\Conid{F}\;(\id{to}\;\Conid{G})\;\Varid{o})\;(\id{iso₁}\;\Conid{F}\;\__{}\;\Varid{o}\;\__{})\;\con{⟩}\;{}\<[E]%
\\[\blanklineskip]%
\>[5]{}\hsindent{3}{}\<[8]%
\>[8]{}\highlight{\id{map}_{\kern0.05em \Varid{i}}\;\Conid{F}\;}(\id{to}\;\Conid{G})\;\Varid{o}\;(\highlight{\id{map}_{\kern0.05em \Varid{i}}\;\Conid{F}\;}(\id{from}\;\Conid{G})\;\Varid{o}\;\Varid{x})\;{}\<[E]%
\\[\blanklineskip]%
\>[3]{}\hsindent{2}{}\<[5]%
\>[5]{}\Varid{≡⟨}\;\Varid{sym}\;(\id{map}^{\mathbin{\circ}}_{\kern0.05em \Varid{i}}\;\Conid{F}\;(\id{to}\;\Conid{G})\;(\id{from}\;\Conid{G})\;\Varid{o}\;\Varid{x})\;\con{⟩}\;{}\<[E]%
\\[\blanklineskip]%
\>[5]{}\hsindent{2}{}\<[7]%
\>[7]{}\id{map}_{\kern0.05em \Varid{i}}\;\Conid{F}\;(\highlight{\id{to}\;\Conid{G}\mathbin{\idsym{∘\ensuremath{_⇉}}_{\kern0.05em \Varid{i}}}\id{from}\;\Conid{G}})\;\Varid{o}\;\Varid{x}\;{}\<[E]%
\\[\blanklineskip]%
\>[3]{}\hsindent{2}{}\<[5]%
\>[5]{}\Varid{≡⟨}\;\id{map}^{\Varid{∀}}_{\kern0.05em \Varid{i}}\;\Conid{F}\;(\id{iso₁}\;\Conid{G}\;\Varid{r})\;\Varid{o}\;\Varid{x}\;\con{⟩}\;{}\<[E]%
\\[\blanklineskip]%
\>[5]{}\hsindent{2}{}\<[7]%
\>[7]{}\highlight{\id{map}_{\kern0.05em \Varid{i}}\;\Conid{F}\;\id{id\ensuremath{_⇉}}_{\kern0.05em \Varid{i}}\;}\Varid{o}\;\Varid{x}\;{}\<[E]%
\\[\blanklineskip]%
\>[3]{}\hsindent{2}{}\<[5]%
\>[5]{}\Varid{≡⟨}\;\id{map}^{\id{id}}_{\kern0.05em \Varid{i}}\;\Conid{F}\;\Varid{o}\;\Varid{x}\;\con{⟩}\;{}\<[E]%
\\[\blanklineskip]%
\>[5]{}\hsindent{2}{}\<[7]%
\>[7]{}\Varid{x}\;\qed{}\<[E]%
\ColumnHook
\end{hscode}\resethooks
The proof for composition is relatively simple, relying on applying the proof
recursively, fusing the two maps, reasoning by recursion on the resulting map,
which results in an identity map. The proof for fixed points is slightly more
involved:

\begin{hscode}\SaveRestoreHook
\column{B}{@{}>{\hspre}l<{\hspost}@{}}%
\column{3}{@{}>{\hspre}l<{\hspost}@{}}%
\column{5}{@{}>{\hspre}l<{\hspost}@{}}%
\column{7}{@{}>{\hspre}l<{\hspost}@{}}%
\column{61}{@{}>{\hspre}l<{\hspost}@{}}%
\column{E}{@{}>{\hspre}l<{\hspost}@{}}%
\>[3]{}\id{iso₁}\;(\con{Fix}_{\kern0.05em \Varid{i}}\;\Conid{F})\;\Varid{r}\;\Varid{o}\;\con{⟨}\Varid{x}\con{⟩}_{\kern0.05em \Varid{i}}\;\mathrel{=}\;\Varid{cong}\;\con{⟨}\_\con{⟩}_{\kern0.05em \Varid{i}}\;\mathbin{\$}\;{}\<[E]%
\\
\>[3]{}\hsindent{2}{}\<[5]%
\>[5]{}\Varid{begin}\;{}\<[E]%
\\[\blanklineskip]%
\>[5]{}\hsindent{2}{}\<[7]%
\>[7]{}\id{map}_{\kern0.05em \Varid{i}}\;\Conid{F}\;(\id{id\ensuremath{_⇉}}_{\kern0.05em \Varid{i}}\mathbin{\idsym{∥}_{\kern0.05em \Varid{i}}}(\id{to}\;(\con{Fix}_{\kern0.05em \Varid{i}}\;\Conid{F})))\;\Varid{o}\;(\highlight{\id{to}\;\Conid{F}\;\Varid{o}\;(\id{from}\;\Conid{F}\;}\Varid{o}\;(\id{map}_{\kern0.05em \Varid{i}}\;\Conid{F}\;(\id{id\ensuremath{_⇉}}_{\kern0.05em \Varid{i}}\mathbin{\idsym{∥}_{\kern0.05em \Varid{i}}}(\id{from}\;(\con{Fix}_{\kern0.05em \Varid{i}}\;\Conid{F})))\;\Varid{o}\;\Varid{x})))\;{}\<[E]%
\\[\blanklineskip]%
\>[3]{}\hsindent{2}{}\<[5]%
\>[5]{}\Varid{≡⟨}\;\Varid{cong}\;(\id{map}_{\kern0.05em \Varid{i}}\;\Conid{F}\;(\id{id\ensuremath{_⇉}}_{\kern0.05em \Varid{i}}\mathbin{\idsym{∥}_{\kern0.05em \Varid{i}}}(\id{to}\;(\con{Fix}_{\kern0.05em \Varid{i}}\;\Conid{F})))\;\Varid{o})\;(\id{iso₁}\;\Conid{F}\;\__{}\;\Varid{o}\;\__{})\;\con{⟩}\;{}\<[E]%
\\[\blanklineskip]%
\>[5]{}\hsindent{2}{}\<[7]%
\>[7]{}\highlight{\id{map}_{\kern0.05em \Varid{i}}\;\Conid{F}\;}(\id{id\ensuremath{_⇉}}_{\kern0.05em \Varid{i}}\mathbin{\idsym{∥}_{\kern0.05em \Varid{i}}}(\id{to}\;(\con{Fix}_{\kern0.05em \Varid{i}}\;\Conid{F})))\;\Varid{o}\;(\highlight{\id{map}_{\kern0.05em \Varid{i}}\;\Conid{F}\;}(\id{id\ensuremath{_⇉}}_{\kern0.05em \Varid{i}}\mathbin{\idsym{∥}_{\kern0.05em \Varid{i}}}(\id{from}\;(\con{Fix}_{\kern0.05em \Varid{i}}\;\Conid{F})))\;\Varid{o}\;\Varid{x})\;{}\<[E]%
\\[\blanklineskip]%
\>[3]{}\hsindent{2}{}\<[5]%
\>[5]{}\Varid{≡⟨}\;\Varid{sym}\;(\id{map}^{\mathbin{\circ}}_{\kern0.05em \Varid{i}}\;\Conid{F}\;(\id{id\ensuremath{_⇉}}_{\kern0.05em \Varid{i}}\mathbin{\idsym{∥}_{\kern0.05em \Varid{i}}}(\id{to}\;{}\<[61]%
\>[61]{}(\con{Fix}_{\kern0.05em \Varid{i}}\;\Conid{F})))\;(\id{id\ensuremath{_⇉}}_{\kern0.05em \Varid{i}}\mathbin{\idsym{∥}_{\kern0.05em \Varid{i}}}(\id{from}\;(\con{Fix}_{\kern0.05em \Varid{i}}\;\Conid{F})))\;\Varid{o}\;\Varid{x})\;\con{⟩}\;{}\<[E]%
\\[\blanklineskip]%
\>[5]{}\hsindent{2}{}\<[7]%
\>[7]{}\id{map}_{\kern0.05em \Varid{i}}\;\Conid{F}\;(\highlight{\id{id\ensuremath{_⇉}}_{\kern0.05em \Varid{i}}\mathbin{\idsym{∥}_{\kern0.05em \Varid{i}}}(\id{to}\;(\con{Fix}_{\kern0.05em \Varid{i}}\;\Conid{F}))})\mathbin{\idsym{∘\ensuremath{_⇉}}_{\kern0.05em \Varid{i}}}(\highlight{\id{id\ensuremath{_⇉}}_{\kern0.05em \Varid{i}}\mathbin{\idsym{∥}_{\kern0.05em \Varid{i}}}(\id{from}\;(\con{Fix}_{\kern0.05em \Varid{i}}\;\Conid{F}))})\;\Varid{o}\;\Varid{x}\;{}\<[E]%
\\[\blanklineskip]%
\>[3]{}\hsindent{2}{}\<[5]%
\>[5]{}\Varid{≡⟨}\;\Varid{sym}\;(\id{map}^{\Varid{∀}}_{\kern0.05em \Varid{i}}\;\Conid{F}\;\Varid{∥∘}_{\kern0.05em \Varid{i}}\;\Varid{o}\;\Varid{x})\;\con{⟩}\;{}\<[E]%
\\[\blanklineskip]%
\>[5]{}\hsindent{2}{}\<[7]%
\>[7]{}\id{map}_{\kern0.05em \Varid{i}}\;\Conid{F}\;((\highlight{\id{id\ensuremath{_⇉}}_{\kern0.05em \Varid{i}}\mathbin{\idsym{∘\ensuremath{_⇉}}_{\kern0.05em \Varid{i}}}\id{id\ensuremath{_⇉}}_{\kern0.05em \Varid{i}}})\mathbin{\idsym{∥}_{\kern0.05em \Varid{i}}}(\highlight{\id{to}\;(\con{Fix}_{\kern0.05em \Varid{i}}\;\Conid{F})\mathbin{\idsym{∘\ensuremath{_⇉}}_{\kern0.05em \Varid{i}}}\id{from}\;(\con{Fix}_{\kern0.05em \Varid{i}}\;\Conid{F})}))\;\Varid{o}\;\Varid{x}\;{}\<[E]%
\\[\blanklineskip]%
\>[3]{}\hsindent{2}{}\<[5]%
\>[5]{}\Varid{≡⟨}\;\id{map}^{\Varid{∀}}_{\kern0.05em \Varid{i}}\;\Conid{F}\;(\id{∥\ensuremath{_\text{cong}}}_{\kern0.05em \Varid{i}}\;(\Varid{λ}\;\__{}\;\__{}\;\Varid{→}\;\con{refl})\;(\id{iso₁}\;(\con{Fix}_{\kern0.05em \Varid{i}}\;\Conid{F})\;\Varid{r}))\;\Varid{o}\;\Varid{x}\;\con{⟩}\;{}\<[E]%
\\[\blanklineskip]%
\>[5]{}\hsindent{2}{}\<[7]%
\>[7]{}\id{map}_{\kern0.05em \Varid{i}}\;\Conid{F}\;(\highlight{\id{id\ensuremath{_⇉}}_{\kern0.05em \Varid{i}}\mathbin{\idsym{∥}_{\kern0.05em \Varid{i}}}\id{id\ensuremath{_⇉}}_{\kern0.05em \Varid{i}}})\;\Varid{o}\;\Varid{x}\;{}\<[E]%
\\[\blanklineskip]%
\>[3]{}\hsindent{2}{}\<[5]%
\>[5]{}\Varid{≡⟨}\;\id{map}^{\Varid{∀}}_{\kern0.05em \Varid{i}}\;\Conid{F}\;(\id{∥\ensuremath{_\text{id}}}_{\kern0.05em \Varid{i}}\;(\Varid{λ}\;\__{}\;\__{}\;\Varid{→}\;\con{refl})\;(\Varid{λ}\;\__{}\;\__{}\;\Varid{→}\;\con{refl}))\;\Varid{o}\;\Varid{x}\;\con{⟩}\;{}\<[E]%
\\[\blanklineskip]%
\>[5]{}\hsindent{2}{}\<[7]%
\>[7]{}\highlight{\id{map}_{\kern0.05em \Varid{i}}\;\Conid{F}\;\id{id\ensuremath{_⇉}}_{\kern0.05em \Varid{i}}\;}\Varid{o}\;\Varid{x}\;{}\<[E]%
\\[\blanklineskip]%
\>[3]{}\hsindent{2}{}\<[5]%
\>[5]{}\Varid{≡⟨}\;\id{map}^{\id{id}}_{\kern0.05em \Varid{i}}\;\Conid{F}\;\Varid{o}\;\Varid{x}\;\con{⟩}\;{}\<[E]%
\\[\blanklineskip]%
\>[5]{}\hsindent{2}{}\<[7]%
\>[7]{}\Varid{x}\;\qed{}\<[E]%
\ColumnHook
\end{hscode}\resethooks
We start in the same way as with composition, but once we have fused the maps
we have to deal with the fact that we are mapping distinct functions to the
left (arguments) and right (recursive positions). We proceed with a lemma on
disjunctive maps that states that a composition of disjunctions is the
disjunction of the compositions (\ensuremath{\Varid{∥∘}_{\kern0.05em \Varid{i}}}). Then we are left with a
composition of identities on the left, which we solve with reflexivity, and a
composition of \ensuremath{\id{to}} and \ensuremath{\id{from}} on the right, which we solve by induction.
Finally, we show that a disjunction of identities is the identity (with the
\ensuremath{\id{∥\ensuremath{_\text{id}}}_{\kern0.05em \Varid{i}}} lemma), and that the identity map is the identity. The lemmas
regarding \ensuremath{\id{map}_{\kern0.05em \Varid{i}}} that we require are the following:

\begin{hscode}\SaveRestoreHook
\column{B}{@{}>{\hspre}l<{\hspost}@{}}%
\column{3}{@{}>{\hspre}l<{\hspost}@{}}%
\column{21}{@{}>{\hspre}l<{\hspost}@{}}%
\column{24}{@{}>{\hspre}l<{\hspost}@{}}%
\column{E}{@{}>{\hspre}l<{\hspost}@{}}%
\>[3]{}\id{map}^{\mathbin{\circ}}_{\kern0.05em \Varid{i}}\;{}\<[21]%
\>[21]{}\mathbin{:}\;{}\<[24]%
\>[24]{}\{\mskip1.5mu \Conid{I}\;\Conid{O}\;\mathbin{:}\;\ty{Set}\mskip1.5mu\}\;\{\mskip1.5mu \Varid{r}\;\Varid{s}\;\Varid{t}\;\mathbin{:}\;\id{Indexed}\;\Conid{I}\mskip1.5mu\}\;{}\<[E]%
\\
\>[24]{}(\Conid{C}\;\mathbin{:}\;\ty{Code}_{\kern0.05em \Varid{i}}\;\Conid{I}\;\Conid{O})\;(\Varid{f}\;\mathbin{:}\;\Varid{s}\mathbin{\idsym{⇉}_{\kern0.05em \Varid{i}}}\Varid{t})\;(\Varid{g}\;\mathbin{:}\;\Varid{r}\mathbin{\idsym{⇉}_{\kern0.05em \Varid{i}}}\Varid{s})\;\Varid{→}\;\id{map}_{\kern0.05em \Varid{i}}\;\Conid{C}\;(\Varid{f}\mathbin{\idsym{∘\ensuremath{_⇉}}_{\kern0.05em \Varid{i}}}\Varid{g})\mathbin{\idsym{≅}_{\kern0.05em \Varid{i}}}(\id{map}_{\kern0.05em \Varid{i}}\;\Conid{C}\;\Varid{f}\mathbin{\idsym{∘\ensuremath{_⇉}}_{\kern0.05em \Varid{i}}}\id{map}_{\kern0.05em \Varid{i}}\;\Conid{C}\;\Varid{g}){}\<[E]%
\\[\blanklineskip]%
\>[3]{}\id{map}^{\Varid{∀}}_{\kern0.05em \Varid{i}}\;{}\<[21]%
\>[21]{}\mathbin{:}\;{}\<[24]%
\>[24]{}\{\mskip1.5mu \Conid{I}\;\Conid{O}\;\mathbin{:}\;\ty{Set}\mskip1.5mu\}\;\{\mskip1.5mu \Varid{r}\;\Varid{s}\;\mathbin{:}\;\id{Indexed}\;\Conid{I}\mskip1.5mu\}\;\{\mskip1.5mu \Varid{f}\;\Varid{g}\;\mathbin{:}\;\Varid{r}\mathbin{\idsym{⇉}_{\kern0.05em \Varid{i}}}\Varid{s}\mskip1.5mu\}\;{}\<[E]%
\\
\>[24]{}(\Conid{C}\;\mathbin{:}\;\ty{Code}_{\kern0.05em \Varid{i}}\;\Conid{I}\;\Conid{O})\;\Varid{→}\;\Varid{f}\mathbin{\idsym{≅}_{\kern0.05em \Varid{i}}}\Varid{g}\;\Varid{→}\;\id{map}_{\kern0.05em \Varid{i}}\;\Conid{C}\;\Varid{f}\mathbin{\idsym{≅}_{\kern0.05em \Varid{i}}}\id{map}_{\kern0.05em \Varid{i}}\;\Conid{C}\;\Varid{g}{}\<[E]%
\\[\blanklineskip]%
\>[3]{}\id{map}^{\id{id}}_{\kern0.05em \Varid{i}}\;{}\<[21]%
\>[21]{}\mathbin{:}\;{}\<[24]%
\>[24]{}\{\mskip1.5mu \Conid{I}\;\Conid{O}\;\mathbin{:}\;\ty{Set}\mskip1.5mu\}\;\{\mskip1.5mu \Varid{r}\;\mathbin{:}\;\id{Indexed}\;\Conid{I}\mskip1.5mu\}\;(\Conid{C}\;\mathbin{:}\;\ty{Code}_{\kern0.05em \Varid{i}}\;\Conid{I}\;\Conid{O})\;\Varid{→}\;\id{map}_{\kern0.05em \Varid{i}}\;\{\mskip1.5mu \Varid{r}\;\mathrel{=}\;\Varid{r}\mskip1.5mu\}\;\Conid{C}\;\id{id\ensuremath{_⇉}}_{\kern0.05em \Varid{i}}\mathbin{\idsym{≅}_{\kern0.05em \Varid{i}}}\id{id\ensuremath{_⇉}}_{\kern0.05em \Varid{i}}{}\<[E]%
\ColumnHook
\end{hscode}\resethooks
Like the lemmas for \ensuremath{\id{map}_{\Varid{r}}} of \autoref{sec:Regular2PolyP}, these are
trivially proven by induction on the structure of \indexed codes.

\section{Discussion and future work}
\label{sec:discussion}%

We have compared different generic programming universes by showing the
inclusion relation between them. This is useful to determine that one approach
can encode at least as many datatypes as another approach, and also allows for
lifting representations between compatible approaches. This also means that
generic functions from approach \text{\tt B} are all applicable in approach \text{\tt A}, if \text{\tt A}
embeds into \text{\tt B}, because we can bring generic values from approach \text{\tt A} into
\text{\tt B} and apply the function there.
However, we cannot make statements
about the variety of generic functions that can be encoded in each approach. The
generic map function, for instance, cannot be defined in \ig, while it is
standard in \indexed.
One possible direction for future research is to
devise a formal framework for evaluating what generic functions are possible
in each universe, adding another dimension to the comparison of the approaches.

Notably absent from our comparison are libraries with a generic view not based
on a sum of products. In particular, the spine view~\cite{SYBreloaded} is
radically different from the approaches we model;
yet, it is the basis for a number of popular generic programming libraries.
It would be interesting to see how these approaches relate to those we have
seen, but, at the moment, converting between entirely different universes
remains a challenge.

An issue that remains with our modelling is to properly address termination.
While our conversion functions can be used operationally to enable portability
across different approaches, to serve as a formal proof they have to be
terminating. Since Agda's algorithm for checking termination is highly
syntax-driven, attempts to convince Agda of termination are likely
to clutter the model, making it less easy to understand. We have thus decided
to postpone such efforts for future work, perhaps relying on sized types for
guaranteeing termination of our proofs~\cite{MiniAgda}.

A related issue that remains to be addressed is
our use of \text{\tt \char45{}\char45{}type\char45{}in\char45{}type} in~\autoref{sec:Indexed2IG}, for the code conversion
function \ensuremath{{}_{\kern0.05em \Varid{i}}\!\!\Uparrow^{_{\Varid{ig}}}\!\!}. It was not immediately clear to us how to solve this issue
even with the recently added support for universe polymorphism in Agda.

Nonetheless, we believe that our work is an important first step towards a
formal categorisation of generic programming libraries. Future approaches
can now rely on our formalisation to describe precisely the new aspects
they introduce, and how the new approach relates to existing ones. In this way
we can hope for a future of \emph{modular generic programming}, where a
specific library might be constructed using components from different
approaches, tailored to a particular need while still reusing code from
existing libraries.

\bibliographystyle{eptcs}
\bibliography{FormalComparisonGP}

\end{document}